\documentclass[twocolumn,preprintnumbers,amsmath,amssymb]{revtex4}
\usepackage{amsmath}
\usepackage{extarrows}
\usepackage[ruled]{algorithm2e}
\usepackage{graphicx}
\usepackage{dcolumn}
\usepackage{bm}
\usepackage[all]{xy}
\usepackage{indentfirst}
\usepackage{amsmath}
\usepackage{multirow}
\usepackage{mathrsfs}
\usepackage{bbm}
\usepackage{euscript}
\usepackage{amssymb}
\usepackage{extarrows}
\usepackage{bbm}
\usepackage[ruled]{algorithm2e}
\usepackage{graphicx}
\begin{document}
\title{Catalytic Entropy Principles}

\author{M.-X. Luo $^{1}$, X. Wang $^2$}

\affiliation{\small{} $^1$ School of Information Science and Technology, Southwest Jiaotong University, Chengdu 610031, China;
\\
\small{}$^2$ School of Electronic Engineering, Dublin City University, Dublin 9, Ireland}

\begin{abstract}
The entropy shows an unavoidable tendency of disorder in thermostatistics according to the second thermodynamics law. This provides a minimization entropy principle for quantum thermostatistics with the von Neumann entropy and nonextensive quantum thermostatistics with special Tsallis entropy. Our goal in this work is to provide operational characterizations of general entropy measures. We present the first catalytic principle consistent with the second thermodynamics law in terms of general quantum entropies for both quantum thermostatistics and nonextensive quantum thermostatistics. This further reveals new features beyond the second thermodynamics law by maximizing the cross-entropy during irreversible catalytic procedures. The present result is useful for asymptotical tasks of quantum entropy estimations and universal quantum source encoding without state tomography. It is further applied to single-shot state transitions and cooling in quantum thermodynamics with limited  information. These results should be interesting in the many-body theory and long-range quantum information processing.
\end{abstract}
\maketitle

The change of entropy of a given system is generally characterized by the second thermodynamics law. By maximizing the entropy in the thermodynamic limit, the thermodynamical distribution of the most probable macrostate under the given energy is determined by the discrete Maxwell-Boltzmann law. This allows an explicit formulation of entropy in terms of Gibbs distribution. For microstates, the thermodynamic entropy is proportional to the logarithm of the number of microstates. Within the Hilbert space of microstates, the entropy is then associated with a statistical operator (mixture of pure states) by von Neumann \cite{Neumann}. The so called von Neumann entropy is formalized as $S(\rho)=-{\rm tr}\rho\log\rho$ for each statistical operator $\rho$. This physical quantity is independent of Schatten decomposition of $\rho$ but its spectra. It also suggests an operational definition of entanglement using wave functions. The von Neumann entropy has been widely used in quantum many-body theory \cite{LP,Eisert,Laflor} and quantum information theory for asymptotic tasks \cite{Schu,BBPS} or single-shot tasks  \cite{MBDR,BHO,Datta,Boes18,Boes19}.

As a parameterized generalization Renyi proposes an information measure preserving the additivity which is compatible with classical probability \cite{Renyi}. It also provides additional insight into the entanglement spectrum such as the largest eigenvalue and number of nonvanishing eigenvalues \cite{Hartley}. Renyi entropy is an important diagnostic probe in the information theory and condensed matter physics. Another extension is Tsallis entropy \cite{Tsallis}. As a nonextensive statistics, it is consistent with Laplace's maximum ignorance principle \cite{Tsallis1} that implies a generalized Bobltzmann-Gibbs statistics for longstanding quasi-stationary states in long-range interacting systems \cite{Tsallis1}. Tsallis entropy follows widely applications in thermodynamics \cite{Tsallis1} and biomedical signal processing \cite{GPT}. Note that the von Neumann entropy \cite{Neumann} not only demonstrates the entropy increase of the second law for any measurement process, but also characterizes the capacity of a quantum communication channel \cite{Holevo}. A natural problem is how to extend these interesting ideas for other entropies \cite{Renyi,Tsallis,AR}.

The quantum entropy even for the von Neumann entropy is generally difficult to estimate faithfully. A directive way is the state tomography that generally requires asymptotic unknown states, especially for continuous-variable states. This provides more information going beyond the quantum entropy. An interesting idea is from thermodynamic properties of Hamiltonians that allow measuring the von Neumann entropy \cite{Men}, Renyi entropy \cite{Emidio}, or Tsallis entropy of specific ground states without probing wave functions. Unfortunately, the lattice-based systems does not allow an explicit procedure for operational tasks in quantum information theory without knowledge of Hamiltonians or fully control over the system.

In this work, we provide an operational characterization of universal quantum entropy for both quantum thermostatistics or nonextensive quantum thermostatistics. Our goal is inspired by von Neumann's original idea \cite{Neumann}. We firstly show the unified entropy principles for any unknown states on separable (finite or infinite dimensional) Hilbert space in terms of the von Neumann entropy \cite{Men}, Renyi entropy \cite{Renyi}, Tsallis entropy \cite{Tsallis} and generalized entropies \cite{Abe,Furuichi,Tsallis2,SM}. Interestingly, it sheds new insight into the second thermodynamics law by increasing the cross-entropy during the irreversible measurement procedure even if for nonextensive quantum thermostatistics with Tsallis entropy \cite{Tsallis}. This also extends the quantum communication theory with the von Neumann entropy \cite{Holevo} to general entropy measures. These results further allow estimating quantum entropies for asymptotical unknown states such as coherent states or 2D spin systems without the state tomography or knowledge of Hamiltonian beyond Jaynes's principle \cite{Jaynes}. Additionally, it further implies universal quantum source encoding with only information of dephased states \cite{Schu,Jozsa,Hayashi}. Finally, it is applicable for single-shot state transition and state cooling in quantum thermodynamics.

\section*{Results}

Consider a quantum system on separable Hilbert space $\mathbb{H}$ (admitting a finite or countable orthogonal basis). A pure state is described by a normalized vector $|\Phi\rangle$ in $\mathbb{H}$. A mixed state is represented by density matrix $\rho$ which is a Hermitian, positive semidefinite statistic operator on $\mathbb{H}$ with unit trace.

For a given density operator $\rho$ on separable Hilbert space $\mathbb{H}_a$, the von Neumann entropy \cite{Neumann} is defined by
\begin{eqnarray}
S(\rho)=-{\rm tr}\rho\log \rho
\label{Neumann}
\end{eqnarray}
The von Neumann entropy provides an entanglement measure, i.e., $S(\rho)>0$ if and only if the system $a$ is entangled with some reference. This is further extended to parameterized entropy measures. One is Renyi entropy \cite{Renyi} given by
\begin{eqnarray}
S_\alpha(\rho)=\frac{1}{1-\alpha}\log{\rm tr}\rho^\alpha, \alpha>0
\label{Renyi}
\end{eqnarray}
The other is Tsallis entropy \cite{Tsallis} given by
\begin{eqnarray}
S_q(\rho)=\frac{1}{1-q}({\rm tr}\rho^q-1), q>0
\label{Tsallis}
\end{eqnarray}
Both are related to the von Neumann entropy as: $\lim_{\alpha\to1}S_\alpha(\rho)=\lim_{q\to1}S_q(\rho)=S(\rho)$. Consider the spectra decomposition of $\rho=\sum_i\lambda_i|\phi_i\rangle\langle \phi_i|$, these quantum entropies are respectively consistent with Shannon entropy \cite{Shannon}, Renyi entropy \cite{Renyi}, and Tsallis entropy \cite{Tsallis} of the distribution $\{\lambda_i\}$.

Given an unknown state $\rho$ on separable Hilbert space $\mathbb{H}_a$, let ${\cal D}$ be a projection measurement according to the orthogonal basis $J=\{|\phi_i\rangle\langle \phi_i|\}$ as
\begin{eqnarray}
{\cal D}_J(\rho)=\sum_{i}q_i|\phi_i\rangle \langle \phi_i|
\label{I1}
\end{eqnarray}
where $q_i={\rm tr}(|\phi_i\rangle\langle \phi_i|\rho)$ defines a distribution from Born rule. This provides a classical description of $\rho$ with the preferred basis $J$. In what follows, define $S({\cal D}_J(\rho))$ as the corresponding classical entropy of the distribution $\{q_i\}$ determined by ${\cal D}_J(\rho)$. One fundamental problem is to characterize general quantum entropies of $\rho$ from the observed state ${\cal D}_J(\rho)$. The problem is firstly solved by von Neumann \cite{Neumann}. This shows a manifestation of the second thermodynamics law in terms of the von Neumann entropy. Our goal is to extend for general entropies with irreversible measurement procedures and beyond.

{\bf Theorem 1}(Entropy principle). \textit{For an unknown state $\rho$ on separable Hilbert space $\mathbb{H}_a$, let $|\Phi\rangle_{ab}$ be any one purification on Hilbert space $\mathbb{H}_a\otimes \mathbb{H}_b$. The quantum entropy $S(\rho)$ satisfies the following equalities:
\begin{eqnarray}
S(\rho)&=&\min_{J_{a}}\{S({\cal D}_{J_a}(\rho))\}
\label{eqn-7}
\\
&=&\min_{J_{a},J_b}\{S({\cal D}_{J_a\otimes J_b}(|\Phi\rangle\langle\Phi|)\}
\label{eqn-6}
\\
&=&
\max_{J_a,J_b}\{I({\cal D}_{J_a\otimes J_b}(|\Phi\rangle\langle\Phi|)\}
\label{eqn-8}
\end{eqnarray}
where $I(\rho_{ab})$ denotes the cross entropy defined by $I(\rho_{ab})=S(\rho_a)+S(\rho_b)-S(\rho_{ab})$.}

\begin{figure}
\begin{center}
\resizebox{120pt}{100pt}{\includegraphics{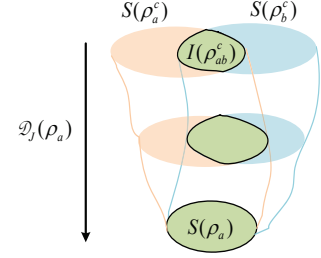}}
\end{center}
\caption{\small (Color online) Schematic entropy principles for quantum measurement procedures. Here, $\rho^c_a$, $\rho^c_b$ and $\rho^c_{ab}$ denotes the observed states on the systems $a$, $b$, and joint system, respectively. The quantum entropy $S(\rho_a)$ increases during any measurement procedures on the local system or purified systems, i.e., $S(\rho_a)\leq \min\{S(\rho^c_a), S(\rho^c_b), S(\rho^c_{ab})\}$. The cross-entropy will decrease during quantum measurement procedures, i.e., $I(\rho_{ab})\geq  I(\rho^c_{ab})$, which shows new insight into the second thermodynamics law.}
\label{fig-1}
\end{figure}

Theorem 1 provides universal consistency with the second law of thermodynamics in both quantum thermostatistics with the von Neumann entropy \cite{Neumann,Renyi} and nonextensive quantum thermostatistics with Tsallis entropy \cite{Tsallis} or generalized entropies \cite{Abe,Furuichi,Tsallis2,SM}. It is going beyond previous results of the von Neumann entropy \cite{Neumann} or Tsallis entropy \cite{Tsallis} with the entropic index $q\in (0,2]$. Especially, Eq.(\ref{eqn-7}) provides a unified demonstration of the second thermodynamics law for any local measurement processes in terms of any operational entropies. Eq.(\ref{eqn-6}) shows an interesting principle for the global system. The quantum entropy of the local system is consistent with the minimal joint entropy of the observed global state, as shown in Fig.\ref{fig-1}. This is from the simultaneous arrangements of two microsystems which implies no additional information being provided by the reference. The idea is further justified by the cross entropy in Eq.(\ref{eqn-8}), which sheds new insights into the second thermodynamics law. The cross entropy presents the capacity of a quantum communication channel inspired by any bipartite entanglement $|\Phi\rangle_{ab}$ in terms of any entropy measures beyond the von Neumann entropy \cite{Holevo}. The proof of Theorem 1 is from the convexity of specific entropy functions. The detailed proofs for the von Neumann entropy \cite{Neumann}, Renyi entropy \cite{Renyi}, Tsallis entropy \cite{Tsallis} and any generalized entropy under proper axioms \cite{Abe,Furuichi,Tsallis2,SM} are respectively presented in Appendixes A-D.

Theorem 1 implies various applications in quantum information processing. The first is a unified uncertainty principle as
\begin{eqnarray}
S({\cal D}_{J_1}(\rho))+S({\cal D}_{J_2}(\rho))\geq 2 S(\rho)
\label{uncert}
\end{eqnarray}
for any two measurement bases $J_1=\{|\phi_i\rangle\}$ and $J_2=\{|\varphi_j\rangle\}$. This provides an optimal bound beyond Maassen-Uffink uncertainty principle \cite{MU} with a lower bound $-2 \log\sup_{i,j}\{\langle \phi_i |\varphi_j\rangle\}$ and Frank-Lieb uncertainty principle \cite{FL} with the lower bound $S(\rho)-2\log c$ in terms of the von Neumann entropy. Another is to demonstrate Araki-Lieb inequality \cite{AL}: $|S(\rho_a)-S(\rho_b)|\leq S(\rho_{ab})$ which may violate Shannon additivity inequality \cite{Shannon,SSI} (Appendix E). Another is to generalize Hadamard-Fischer determinant inequality \cite{Bhatia} into the unified entropy inequality on separable Hilbert space as
\begin{eqnarray}
S(\rho)\leq S({\cal D}_{J}(\rho))
\label{HDI}
\end{eqnarray}
where $J$ denotes any orthogonal basis.

\textbf{Entropy estimation of unknown states}. Consider a quantum source generating an ensemble orthogonal states $|\phi_i\rangle$ under the distribution $\{p_i\}$, i.e., $\rho=\sum_{i=1}^np_i|\phi_i\rangle\langle \phi_i|$. In the experiment, the local system of an unideal many-body source may be decohered quickly in some given basis, i.e., $\rho$ being dephased to ${\cal D}_J(\rho)=\sum_{j}q_j|\psi_j\rangle\langle \psi_j|$. With this assumption, the goal is to estimate the quantum entropy $S(\rho)$ from the asymptotic resources of ${\cal D}_J(\rho)$. The task generally requires state tomography or Hamiltonians \cite{Men,Emidio}. It seems that the dephased state provides incomplete information for the initial state. Remarkably, Theorem 1 implies an information optimization for estimating $S(\rho)$ assisted by ${\cal D}_J(\rho)$. It will be regarded as a partially statistical inference \cite{Neumann,Jaynes} for the goal as:
\begin{eqnarray}
&&\mathrm{argmax}\quad S(\rho)
\nonumber\\
&&\mathrm{s.t.}, \sum_{i=1}^np_i\alpha_{ij}=q_j, \forall j,
\nonumber\\
 &&\qquad\sum_{i}p_i=1,
\nonumber\\
 &&\qquad\alpha_{ij}=|\langle \phi_i|\psi_j\rangle|^2, \forall i, j.
 \label{argmax}
\end{eqnarray}
Here, $\{q_j\}$ is an observed distribution. Although $\{q_j\}$ and $\{\alpha_{ij}\}$ provides only partial information, the real system behaves maximally noncommittal about missing information \cite{Neumann,Jaynes}. The present method may be interpreted as the maximal information will be recovered by a given irreversible experiment. In applications, the maximal entropy is optimized over all the complement basis states when the partial basis is known. Another is from $m<n$ due to losing information. Interestingly, these entropies are analytically represented by a partition function using the Lagrange multiplier method \cite{Jaynes,Jaynes2} (Appendix F).

\textbf{Quantum unknown source encoding}. For an independent and identically distributed (i.i.d.) source, Shannon Theorem \cite{Shannon} characterizes the redundancy information with a fundamental limit that is achievable for a noiseless channel. The main idea is from the asymptotic equipartition property of typical series, i.e., the joint distribution of typical sequences is asymptotically dominated by its Shannon entropy. A similar result holds for quantum sources \cite{Schu} in terms of the von Neumann entropy \cite{Neumann} by using typical states. Here, inspired by Theorem 1 we provide another method to compress an unknown quantum source with only partial information of the measured state $\rho^c={\cal D}_J(\rho)$ \cite{Neumann}.

{\bf Theorem 2}. \textit{Let $\{\rho_n=\rho^{\otimes n}, \mathbb{H}_n=\mathbb{H}^{\otimes n}\}$ be an i.i.d unknown quantum source. If $R>S(\rho^c)$ with the von Neumann entropy $S(\rho^c)$, there exists a reliable compression scheme of rate $R$ for $\rho_n$}.

In information theory, there is a universal typical set \cite{Cziszar} for any probability distribution with a given Shannon entropy \cite{Shannon}. A similar result holds for quantum sources with the von Neumann entropy \cite{Jozsa,Neumann}, which relies on a typical subspace of all input states. Another quantum variable-length code \cite{Hayashi} depends on the trade-off between the von Neumann entropy and the non-demolition measurement of all input states. Compared with these universal schemes \cite{Jozsa,Hayashi}, Theorem 2 depends only on the von Neumann entropy $S(\rho^c)$ instead of $S(\rho)$. This costs a small fraction of input states while all the remained systems will be universally compressed. The new scheme provides a weak universal quantum compression for unknown sources. The proof of Theorem 2 is shown in Appendix G.

\textbf{One-shot unknown state transition}. How to feature the one-shot state transition from $\rho$ to $\rho'$ on separable Hilbert space $\mathbb{H}_a$ is an important problem in quantum information processing. There are lots of results related to specific conditions \cite{Schu,BBPS,Boes18,Boes19,RW} that generally require the state tomography or quantum entropy $S(\rho)$. A further problem is to transform an unknown state $\rho$ into a given state with limited prior information. For one-shot scenarios, the unknown state $\rho$ may be catalyzed into $\rho^c$ according to the dephasing channel \cite{Boes18,Boes19} defined in Eq.(\ref{I1}). Our goal here is to complete the state transition $\rho\to \rho'$ with the knowledge of $\rho^c$ and $\rho'$. This is formally featured by the quantum entropy as follows (Appendix H).

\begin{figure}
\begin{center}
\resizebox{200pt}{140pt}{\includegraphics{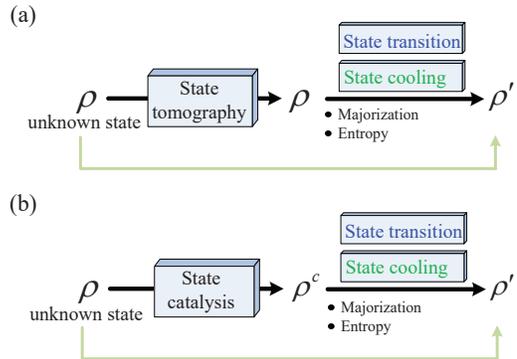}}
\end{center}
\caption{\small (Color online) State transitions under different settings. One is from state tomography. The other is using the correlated catalyst under any dephasing procedure. State transitions in both cases can be featured by using the majorization and quantum entropy.}
\label{fig-2}
\end{figure}

\textbf{Theorem 3}. \textit{If $S(\rho')>S(\rho^c)$ and ${\rm rank}(\rho')\geq {\rm rank}(\rho^c)$, there exist a density matrix $\sigma_b$ on Hilbert space $\mathbb{H}_b$ and a unitary $U$ such that
\begin{eqnarray}
&&{\rm tr}_b[U(\rho\otimes \sigma)U^\dag]=\rho',
\\
&&{\cal D}_J[{\rm tr}_a(U(\rho\otimes \sigma)U^\dag)]=\sigma
\label{eqn42}
\end{eqnarray}
where $S(\rho)$ denotes the von Neumann entropy.}

Theorem 3 also holds for other entropies \cite{Renyi,Tsallis} from their transition relationships. Similar result may be proved for the majorization condition (Appendix H) \cite{Gour}, approximate transition of infinite-dimensional states (Appendix I) while recent result is used for approximate transition of finite states \cite{Wil}, or probabilistic transition of any states (Appendix J). So far, all the results \cite{Gour,Boes19,RW,Wil} are for the known state $\rho$. Theorem 3 implies a sufficient condition to complete state transition without the state tomography, as shown in Fig.\ref{fig-2}. This is reasonable from Theorem 1, i.e., $S(\rho)\leq S(\rho')$ if $S(\rho^c)\leq S(\rho')$. A direct application of Theorem 3 is to get an approximate catalytic state, i.e., $D(\rho',{\rm tr}_b[U(\rho\otimes \mathbf{1})U^\dag])<\epsilon$ with the maximally mixed state $\mathbf{1}$ and any negligible constant $\epsilon>0$ if $D(\rho,\rho^c)<\epsilon$ for any contractive metric $D(\cdot,\cdot)$ which is the unitary invariant. This is applicable for cooling an unknown thermal state into an approximate pure state \cite{BBPS,Boes19}.

\textbf{Quantum entropy of Gaussian states}. Consider an $n$-mode zero-mean Gaussian state \cite{Serafini} as $\rho=\frac{1}{C}\exp(-\frac{1}{2}\vec{x}^T\mathbf{H}\vec{x})$, where $C$ is the normalization constant, $\vec{x}$ is the vector of quadrature operators, and $\mathbf{H}$ is a real positive-definite Hamiltonian matrix. Its quantum entropy is invariant under local unitary operations. From Williamson decomposition theorem \cite{Williamson}, $\rho$ is rewritten into $n$ thermal states of $\otimes_{i=1}^n\varrho$ under a unitary operator, where the mean photon number for $\varrho$ depends on the symplectic eigenvalue of its covariance matrix. Suppose that $\rho$ is undergoing an unknown unitary operation such as the degraded Gaussian broadcast channel \cite{Guha}, which transforms one local system $\rho_a$ into a joint state of $\rho_{ab}=U_\lambda(\rho_a|0\rangle\langle 0|)U^\dag_\lambda$ by using the beamsplitter operator $U_\lambda$. The goal is to estimate the quantum entropy of the reduced density operator $\varrho_a={\rm tr}_b(U\rho_{ab}U^\dag)$. Generally, consider the decomposition of $\varrho=\sum_{n=0}^\infty{}p_n|n\rangle\langle n|$ \cite{Williamson}. The quantum entropy depends only on the average energy of the Gaussian thermal state. In the experiment, one can only obtain a measured state $\hat{\varrho}^c=\sum_{n=0}^Nq_n|n\rangle\langle n|$ associated with the first $N+1$ finite Fock states $\{|n\rangle\langle n|\}_{n=0}^{N}$.  From the additivity of quantum entropy (Supplementary D), it implies that $S(\hat{\varrho}^c)\leq S(\rho_a)$ while $\lim_{N\to\infty } S(\hat{\varrho}^c)=S(\rho_a)$. This allows an optimization similar to Eq.(\ref{argmax}) for estimating $S(\rho_a)$ using the projected state $\hat{\varrho}^c$.

\begin{figure}
\begin{center}
\resizebox{240pt}{120pt}{\includegraphics{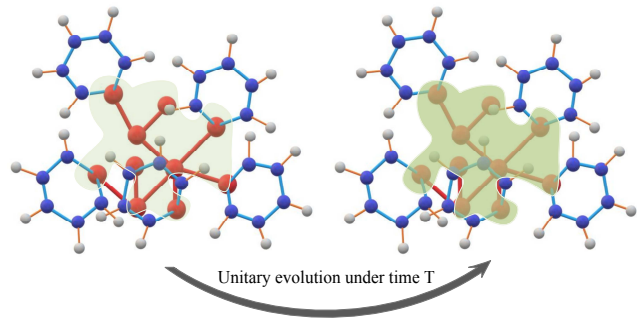}}
\end{center}
\caption{\small (Color online) Schematic illustration of center cluster
(red spins) and surrounding spins (blues). The entropy of the center cluster is changed during joint control of spins under the evolution period $T$. It may be evaluated by quantum dephasing operation.}
\label{fig-3}
\end{figure}

\textbf{2D Spin model}. Consider a 2D center-cluster spin model, as shown in Fig.\ref{fig-3}. The center clusters of spins are coupled to the outer spins according to the interaction as $\mathbf{H}_{cc}=\sum_{s=1}^m\mathbf{H}_{s}$, where $\mathbf{H}_s=\sum_{s=1}^m\sum_{j=1}^n\omega_{sj}Z_s\otimes{}Z_j\otimes\mathbbm{1}_{2^{n+m-2}}$  denote the interactions between the spins in the $s$-th cluster and outer spins, $\mathbbm{1}_{2^{n+m-2}}$ denotes the interactions among the center spins, and $Z_j$ is Pauli matrix $Z$ for the $j$-th spin. Special examples include Ph3P of triphenylphosphine molecules, PCN of tris(cyanoethyl)phosphine or KHB(pz)3 ligands \cite{BGW}. The coupling constants $\omega_{ks}$ are determined by the orientation and the distance of two interacted spins \cite{NSC,NSC1}. Suppose that the central cluster is initially in an unknown state $\rho_{c}(0)$, while the outer spins are in the maximally mixed state $\rho_s(0)= \mathbf{1}_{2^n}$. The control on the external environments \cite{RCC} allows probing multi-spin correlations from the central-cluster spins under the unitary evolution $U_T=\exp(-i\mathbf{H}_{cc}T)$, which yields to the final state $\rho(T)= U_T\rho(0)U_T^\dag$ under the evolution period $T$. Theorem 1 and the optimization (\ref{argmax}) allow estimating the quantum entropy of the center cluster as $S(\rho(T)^c)$ using quantum dephasing operation ${\cal D}_{J}$. One specific example is initialized by the polarized state $\rho(0)=\frac{\mathbbm{1}+X^{\otimes m}}{2^m}$, where $T$ is resolved using the free decay of ${\rm tr}(X^{\otimes m}\rho(T))$. This can be used to approximately evaluate $S(\rho(T))$.

\section*{Discussions}

In a many-body system, the von Neumann entropy \cite{Neumann} quantifies the changes of information from one local system compared with other correlated systems. Instead, the parameterized entropies \cite{Renyi,Tsallis} show additional intuitions for distributing the input information among the composite system during the dynamical evolutions. These demonstrate interesting long-range perspectives for detecting the growth of many-body correlations, the spread of quantum information in composing systems, or asymptotic features in nonextensive quantum thermostatistics. Note that all the parameterized entropies \cite{Renyi,Tsallis} lead to the von Neumann entropy \cite{Neumann} that is generally specified for most one-shot tasks. A fundamental problem is to explore the common features of entropies beyond the second thermodynamics law. The present entropy principles show interesting solutions from unified models for asymptotic tasks or one-shot tasks. This further inspires a twin-class problem of finding distinguished applications of various entropies.

In summary, we provided an operational characterization of universal quantum entropy. The main idea is inspired by the second thermodynamics law for the measurement process. This allows minimal entropy principles for the local system and its purified system during the irreversible measurement. A different feature holds for the maximal cross entropy which provides a unified capacity for quantum communication in terms of various different entropy measures. The present results imply a useful method for the asymptotic task of estimating quantum entropies or universal quantum source encoding. It is further applied for single-shot state transition and cooling in quantum thermodynamics. These results may be applied in entanglement theory, quantum information processing, quantum commutation  or long-range many-body states such as quantum error correction codes.

\section*{Acknowledgements}

We thank Ronald de Wolf, Carlos Palazuelos, Luming Duan, Yaoyun Shi, and Donglin Deng. This work was supported by the National Natural Science Foundation of China (No.61772437), Fundamental Research Funds for the Central Universities (No.2682014CX095), Chuying Fellowship, CSC Scholarship, and EU ICT COST CryptoAction (No.IC1306).

\appendix 

\section{Proof of von Neumann entropy}

We firstly present some basic results for the proof of entropy principles on finite-dimensional Hilbert space or separable Hilbert space.

Let $C\subset{}\mathbb{R}^d$ be a convex bounded set, and $f:C\to \mathbb{R}$ be a convex function. From Jensen inequality \cite{Jensen}, it follows that
\begin{eqnarray}
f(\sum_{i=1}^n\lambda_ix_i)\leq \sum_{i=1}^n\lambda_if(x_i)
\label{A1}
\end{eqnarray}
where $\{\lambda_i\}$ is a distribution, i.e., $\lambda_i>0$ and $\sum_i\lambda_i=1$. Similarly, define an infinite convex combination of elements of $C$ as
\begin{eqnarray}
x_0=\sum_{i=1}^\infty\lambda_ix_i
\label{A2}
\end{eqnarray}
where $x_i\in C, \lambda_i\geq0, \sum_{i=1}^\infty\lambda_i=1$. This Jensen inequality can be extended for infinite-dimensional spaces.

\textbf{Theorem S1} (Jensen inequality) \cite{Jensen}. \textit{Let $(\Omega, \sum,\mu)$ be a probability measure space, and $g:\Omega\to \mathbb{R}^d$ a measurable mapping that is $\mu$-integrable. Let $C\subset\mathbb{R}^d$ be a convex set such that $g(\omega)\in C$ for $\omega\in \Omega$, and $f:C\to \mathbb{R}$ a l.s.c. convex function. Then we have $\int_{\Omega}gd\mu\in C$ and
\begin{eqnarray}
f(\int_\Omega{}gd\mu)\leq \int_{\Omega}f\circ{}gd\mu
\label{A4}
\end{eqnarray}
where l.s.c. stands for lower semicontinuous.
}

There are lots of examples for Theorem S1. We consider the following examples for our goals of entropies as
\begin{itemize}
\item[(i)] Consider $C=(0,1]$ and $f(x)=x\log{}x$. From Theorem S2, it follows that
    \begin{eqnarray}
\int_{\Omega}g\log{}gd\mu\leq \int_{\Omega}g d\mu\log\int_{\Omega}g d\mu
    \label{A3}
    \end{eqnarray}
\item[(ii)] Consider $C=[0,1]$ and $f(x)=x^p(g\geq 1)$. From Theorem S2, it follows that
    \begin{eqnarray}
    \int_{\Omega}gd\mu\leq (\int_{\Omega}g^p)^{1/p}
    \label{A4}
    \end{eqnarray}
\item[(iii)] Consider $C=[0,1]$ and $f(x)=-\log{}x$. From Theorem S2, it follows that
    \begin{eqnarray}
    \int_{\Omega}\log{}gd\mu\leq \log\int_{\Omega}gd\mu
    \label{A5}
    \end{eqnarray}
\end{itemize}

\textbf{Definition S1}. $f:\mathbb{R}^+\to \mathbb{R}$ is $f$-divergence if it satisfies the following properties
\begin{itemize}
\item[(i)] $f$ is a convex function;
\item[(ii)]  $f(1)=0$;
\item[(iii)] $f$ is strictly convex at $x=1$, i.e., $f(1)<\lambda f(x)+(1-\lambda)f(y)$ for all $\lambda\in (0,1), x, y\in \mathbb{R}^+$.
\end{itemize}

Then the functional that maps pairs of distributions to $\mathbb{R}^+$ defined by
\begin{eqnarray}
D_f(P||Q):=\mathbb{E}_Q[f(\frac{dP}{dQ})]
\label{A6}
\end{eqnarray}
is an $f$-divergence, where $\frac{dP}{dQ}$ denotes the Radon-Nikodym derivative \cite{DPI} (or density) of $P$ with respect to $Q$.

\textbf{Theorem S2} (Data Processing Inequality) \cite{DPI}. \textit{Consider a channel that produces $Y$ given $X$ based on the law $P_{Y|X}$ (shown below). If $P_Y$ is the distribution of $Y$ when $X$ is generated by $P_X$ and $Q_Y$ is the distribution of $Y$ when $X$ is generated by $Q_X$, then for any $f$-divergence $D_f(\cdot{}\|\cdot{})$,
\begin{eqnarray}
D_f(P_Y\|Q_Y)\leq D_f(P_X\|Q_X)
\label{A7}
\end{eqnarray}}

Theorem S2 has lots of applications with different formulations of $f$. For our goals, we consider the following definitions.
\begin{itemize}
\item[(i)] \textit{Relative entropy}. For Shannon relative entropy, the function $f$ is defined by $f(t)=-t\log{}t$, which is also Kullback-Leibler divergence \cite{Shannon}.
\item[(ii)] \textit{Renyi divergence}. For the Renyi divergence \cite{Renyi} of order $\alpha\geq0$ from $P$ to $Q$, the function $f$ is defined by $\exp[(\alpha-1)f(t)]=\exp(\alpha{}t)$.
\item[(iii)] \textit{Hellinger divergence}. For Hellinger divergence of Tsallis entropy \cite{Hellinger,Tsallis}, the function $f$ is defined by $f_\alpha=\frac{t^\alpha-1}{\alpha-1}$ with $\alpha\in (0,1)\cup (1,\infty)$.
\end{itemize}

For a general case, consider two separable Hilbert spaces $\mathbb{H}_a$ and $\mathbb{H}_b$. Consider a non-negative bounded Hilbert-Schmidt operator $\rho$ on $\mathbb{H}_x$. From the spectra decomposition theorem, there exist orthogonal basis $\{|f_i\rangle\}$ and spectra $\{\lambda_i\}$ with $\lambda_i\in (0,1)$ and $\sum_{i}\lambda_i=1$, such that
\begin{eqnarray}
\rho=\sum_{i}\lambda_i|f_i\rangle\langle f_i|
\label{A31}
\end{eqnarray}
It is easy to check that ${\rm tr}\rho=\sum_i\lambda_i=1$.

Define $\mathbb{SU}(\mathbb{H}_a\to \mathbb{H}_b)$ consisting of all isometric operators from $\mathbb{H}_a$ to $\mathbb{H}_b$. For any ${\cal U}\in \mathbb{SU}(\mathbb{H}_a\to \mathbb{H}_b)$, define a bounded linear operator $\tilde{\rho}$ on $\mathbb{H}_b$ as \cite{Bassi}:
\begin{eqnarray}
\tilde{\rho}={\cal U}\rho{\cal U}^*=\sum_{i}\lambda_i|\tilde{f}_i\rangle\langle \tilde{f}_i|
\label{A32}
\end{eqnarray}
with $|\tilde{f}_i\rangle={\cal U}|f_i\rangle$. Here, $\tilde{\rho}$ is positive semidefinite Hilbert-Schmidt operator with ${\rm tr}\tilde{\rho}=1$.

For any density operator $\rho=\sum_{i}\lambda_i|f_i\rangle\langle f_i|$ on $\mathbb{H}_a$, similar to purification of finite space, there exists an axillary space $\mathbb{H}_b$ and a rank-1 operator $|F\rangle_{ab}\langle F|$ on $\mathbb{H}_a\otimes \mathbb{H}_b$ such that \cite{Bassi}:
\begin{eqnarray}
{\rm tr}_b(|F\rangle_{ab}\langle F|)=\zeta
\label{A33}
\end{eqnarray}
where ${\rm tr}_b$ denotes the partial trace operator, and $|F\rangle_{ab}$ is represented by
\begin{eqnarray}
|F\rangle_{ab}=\sum_{i}\sqrt{\lambda_i}|f_i\rangle|g_i\rangle
\label{A34}
\end{eqnarray}
and $\{|g_i\rangle\}$ are orthogonal functions on $\mathbb{H}_b$. The rank-1 operator $|F\rangle_{ab}\langle F|$ is named as the purification of $\rho$.

Consider a given system with unknown operator $\rho$ on $\mathbb{H}_a$. Let ${\cal D}$ be the quantum channel that dephases the operator $\rho$ in a given orthogonal basis $J:=\{|g_j\rangle\}$ on $\mathbb{H}_a$ into the following operator
\begin{eqnarray}
{\cal D}_J(\rho)=\sum_{i}q_i|g_i\rangle \langle g_i|
\label{A35}
\end{eqnarray}
where $q_i$ is defined by
\begin{eqnarray}
q_i=\sum_j\lambda_j |\langle g_i|f_j\rangle|^2
\label{A36}
\end{eqnarray}
which defines a distribution on $X$. This provides an observed density operator of $\rho$. Let $X$ be a classical random variable associated with the distribution $\{q_i\}$.  Define von Neumann entropy \cite{Neumann} of density operator $\rho$ as
\begin{eqnarray}
S(\rho)&=&-{\rm tr}[\rho\log \rho]
\nonumber\\
&=&-\sum_i\lambda_i\log\lambda_i
\label{A37}
\end{eqnarray}
from the decomposition in Eq.(\ref{A31}). Here, $S(\rho)=+\infty$ for some operator $\rho$.

Define von Neumann entropy of the density operator ${\cal D}_J(\rho)$ as
\begin{eqnarray}
S({\cal D}_J(\rho))&=&-{\rm tr}[{\cal D}_J(\rho)\log {\cal D}_J(\rho)]
\nonumber\\
&=&-\sum_iq_i\log{}q_i
\label{A38}
\end{eqnarray}
from the decomposition in Eq.(\ref{A35}). Here, $S({\cal D}_J(\rho))$  may be $+\infty$ for some quantum dephasing channels. Our goal in this subsection is to prove the entropy principle for unknown states on separable space $\mathbb{H}_a$.

Now, consider an operator
\begin{eqnarray}
\rho_{ab}=\sum_{i,j}\lambda_{ij}|f_{ij}\rangle\langle f_{ij}|
\label{A39}
\end{eqnarray}
on the space $\mathbb{H}_a\otimes \mathbb{H}_b$, where $\{|f_{ij}\rangle\}$ are orthogonal functions of $\mathbb{H}_a\otimes \mathbb{H}_b$. Define its partial trace operators $\rho_a$ and $\rho_b$ as
\begin{eqnarray}
\rho_a=\sum_{i,j}\lambda_{ij}\langle{}e_i'|f_{ij}\rangle\langle f_{ij}|e_i'\rangle
\\
\rho_b=\sum_{i,j}\lambda_{ij} \langle{}e_i|f_{ij}\rangle\langle f_{ij}|e_i\rangle
\label{A40}
\end{eqnarray}
where $\{|e_i\rangle\}$ and $\{|e_i'\rangle\}$ are complete orthogonal bases of $\mathbb{H}_a$ and $\mathbb{H}_b$, respectively. Since $\mathbb{H}_a$ and $\mathbb{H}_b$ are complete and separable Hilbert spaces, $\{|e_i\rangle|e_i'\rangle\}$ is complete orthogonal basis of $\mathbb{H}_a\otimes \mathbb{H}_b$. It means that $\{|f_{ij}\rangle\}$ is unitarily equivalent to $\{|e_i\rangle|e_i'\rangle\}$. Hence, $\rho_a$ and $\rho_b$ are independent of $\{|e_i'\rangle\}$ and $\{|e_i\rangle\}$, respectively. Here, $\{Pr(x=i,y=j)=\lambda_{ij}\}$ defines a joint distribution of two random variables $x, y$.

Define its conditional entropies $S(\rho_{a|b})$ and $S(\rho_{b|a})$ as
\begin{eqnarray}
&&S(\rho_{a|b})=S(\rho_{ab})-S(\rho_{b})
\label{A41}
\\
&&S(\rho_{b|a})=S(\rho_{ab})-S(\rho_{a})
\label{A42}
\end{eqnarray}
if $\max\{S(\rho_{ab}),S(\rho_{a}),S(\rho_{b})\}<+\infty$. It means that the uncertainties are remained under recovering the partial systems. Here, we do not define conditional density from the Bayesian rule for avoiding $\infty$ in integral. By using these entropies, we can define the mutual information as
\begin{eqnarray}
I(\rho_{ab})&=&S(\rho_{a})-S(\rho_{a|b})
\nonumber\\
&=&S(\rho_{b})-S(\rho_{b|a})
\nonumber\\
&=&S(\rho_{a})+S(\rho_{b})-S(\rho_{ab})
\label{A43}
\end{eqnarray}
This definition is reasonable from Eqs.(\ref{A41}) and (\ref{A42}).

{\bf Theorem S3}(von Neumann Entropy Principle). \textit{For an unknown trace-class operator $\rho$ on separable Hilbert space $\mathbb{H}_a$ and any purification $|F\rangle_{ab}$ on separable Hilbert space $\mathbb{H}_a\otimes \mathbb{H}_b$. Then the following statements hold:
\begin{eqnarray}
S(\rho)&=&\min_{J}\{S({\cal D}_J(\rho))\}
\label{A44}
\\
&=&\min_{J_a,J_b}\{S({\cal D}_{J_1\otimes J_2}(|F\rangle\langle{}F|))\}
\label{A45}
\\
&=&\max_{J_a,J_b}\{I({\cal D}_{J_1\otimes J_2}(|F\rangle\langle{}F|))\}
\label{A46}
\end{eqnarray}
where $S({\cal D}_{J_1\otimes J_2}(|F\rangle\langle{}F|))$ and $I({\cal D}_{J_1\otimes J_2}(|F\rangle\langle{}F|))$ are defined according the joint observed state ${\cal D}_{J_1\otimes J_2}(|F\rangle\langle{}F|)$.
}

{\bf Proof}. It only needs to prove the result for $S(\rho)<+\infty$. Otherwise, it follows from spectra decomposition of $\rho$, i.e., $S({\cal D}_J(\rho))=+\infty$. Eq.(\ref{A44}) is proved by von Neumann \cite{Neumann}. The other proofs are inspired by the von Neumann's idea \cite{Neumann}. From Eq.(\ref{A34}), each purification of the operator $\rho=\sum_i\lambda_i|f_i\rangle\langle f_i|$ on $\mathbb{H}_a$ can be represented by a rank-1 operator $|F\rangle_{ab}\langle F|$ on the space $\mathbb{H}_a\otimes \mathbb{H}_b$ with the function
\begin{eqnarray}
|F\rangle_{ab}=\sum_{i}\sqrt{\lambda_i}|f_i\rangle|g_i\rangle
\label{A47}
\end{eqnarray}
where $\{|g_i\rangle\}$ are orthogonal functions on axillary space $\mathbb{H}_b$. Note that
\begin{eqnarray}
{\rm Pr}(f_i,g_j)
&=&\sum_{i,j}|\langle{}f_i|\langle g_j |F\rangle|^2
=\lambda_i
\label{A48}
\end{eqnarray}
Hence, there is a joint distribution
\begin{eqnarray}
\mathcal{P}_q=\{{\rm Pr}[x=f_i,y=g_j]=\lambda_i\}
\label{A49}
\end{eqnarray}
associated with the function $|F\rangle\langle F|$ by performing rank-1 measurements $\{|f_i\rangle\langle f_i|\}$ and $\{|g_j\rangle\langle g_j|\}$.

The rank-1 function of $|F\rangle_{ab}$ can be changed into
\begin{eqnarray}
|F_0\rangle_{ab}=\sum_{i}\sqrt{\lambda_i}|e_i\rangle_{a}|e'_i\rangle_{b}
\label{A50}
\end{eqnarray}
under the isometric operators ${\cal U}: |f_i\rangle_a\mapsto |e_i\rangle_a$, and ${\cal V}: |g_i\rangle_b\mapsto |e_i'\rangle_b$, where $\{|e_i\rangle\}$ are orthogonal basis on $\mathbb{H}_a$ and $\{|e_i'\rangle\}$ are orthogonal basis on $\mathbb{H}_b$. It follows that by using local operators ${\cal U}$ and ${\cal V}$, the joint distribution $\mathcal{P}_q$ in Eq.(\ref{A49}) is transformed into a new joint distribution
\begin{eqnarray}
\{{\rm Pr}(x_0=e_i,y_0=e_i')=\lambda_i\}
\label{A51}
\end{eqnarray}
associated with $|F_0\rangle\langle F_0|$ under the local rank-1 measurements of $\{|e_i\rangle\langle e_i|\}$ and $\{|e_i'\rangle\langle e_i'|\}$, i.e.,
\begin{eqnarray}
{\rm Pr}(x_0=e_i,y_0=e_i')
&=&\sum_{i,j}|\langle{}e_i e_j' |F_0\rangle|^2
=\lambda_i
\label{A52}
\end{eqnarray}
from Eq.(\ref{A50}), where we have used the following equalities
\begin{eqnarray}
|\langle{}e_ie_j' |e_{s}e_t'\rangle|^2
&=&|\langle{}e_i|e_s\rangle|^2 |\langle{}e_j |e_{t}'\rangle|^2
=\delta_{i,s}\delta_{j,t}
\end{eqnarray}
with the delta function $\delta_{ij}$.

Let $\hat{J}_a=\{|f_i\rangle\}$, $\hat{J}_b=\{|g_i\rangle\}$, $J_a=\{|e_i\rangle\}$ and $J_b=\{|e_i'\rangle\}$. From Eqs.(\ref{A48})-(\ref{A52}), we get
\begin{eqnarray}
S({\cal D}_{\hat{J}_a\otimes \hat{J}_b}(|F\rangle\langle{}F|))
 &=&S({\cal D}_{J_a\otimes J_b}(|F_0\rangle\langle F_0|))
\nonumber\\
&=&H(x_0,y_0)
\nonumber\\
&=&-\sum_{i}\lambda_i\log \lambda_i
\nonumber\\
&=&H(x_0)
\nonumber\\
&=&S(\rho)
\label{A53}
\end{eqnarray}
where $H(x_0)$ denotes the Shannon entropy and $H(x_0,y_0)$ denotes Shannon joint entropy associated with the distribution in Eq.(\ref{A51}), and $S(\rho)$ denotes the von Neumann entropy of the operator $\rho$.

Our goal in what follows is to prove that
\begin{eqnarray}
S({\cal D}_{J_a\otimes J_b}(|F_0\rangle\langle{}F_0|))=\min_{\hat{J}_a\otimes \hat{J}_b}\{S({\cal D}_{\hat{J}_a\otimes \hat{J}_b}(|F\rangle\langle{}F|))\}
\label{A54}
\end{eqnarray}
where $\hat{J}_a$ and $\hat{J}_b$ denote any orthogonal basis functions $\mathbb{H}_a$ and $\mathbb{H}_b$, respectively.

Consider any operators ${\cal U}\in \mathbb{SU}(\mathbb{H}_a)$ and ${\cal V}\in \mathbb{SU}(\mathbb{H}_b)$. Assume that ${\cal U}$ and ${\cal V}$ are respectively defined by
\begin{eqnarray}
{\cal U}: |e_i\rangle \mapsto \sum_{k}\alpha_{ki}|f_k\rangle,  \forall i
\label{A55}
\\
{\cal V}: |e_i'\rangle \mapsto \sum_{k}\beta_{ki}|g_k\rangle, \forall i
\label{A56}
\end{eqnarray}
where $\{|e_i\rangle\}$ and $\{|e_i'\rangle\}$ are complete orthogonal basis functions of $\mathbb{H}_a$ and $\mathbb{H}_b$, respectively, $\alpha_{ki}$ and $\beta_{ki}$ are nonnegative constants satisfying $\alpha_{ki}=|\langle e_i|f_k\rangle|^2$ and $\beta_{ki}=\langle e_i'|g_k\rangle|^2$.

From Eqs.(\ref{A55}) and (\ref{A56}), we get
\begin{eqnarray}
|F_f\rangle=({\cal U}\otimes {\cal V})|F_0\rangle=\sum_{ij}\gamma_{ij}|e_ie_j'\rangle
\label{A57}
\end{eqnarray}
with $\gamma_{ij}=\sum_{k}\sqrt{\lambda_i}\alpha_{ki}\beta_{kj}, \forall i,j$. Hence, we get a distribution associated with the function $|F_f\rangle$ under the local projection measurement with the orthogonal basis function $\{|e_ie_j'\rangle\}$  as
\begin{eqnarray}
\mathcal{P}_q=\{{\rm Pr}[X_f=e_i,Y_f=e_j']=|\gamma_{ij}|^2\}
\label{A58}
\end{eqnarray}
where ${\rm Pr}(X_f=e_i,Y_f=e_j')$ is evaluated as
\begin{eqnarray}
{\rm Pr}(x_0=e_i,y_0=e_i')
&=&\sum_{i,j}|\langle{}e_ie_j'|F_f\rangle|^2
=|\gamma_{ij}|^2
\end{eqnarray}

Note that
\begin{eqnarray}
{\rm Pr}[x_f=e_i]&=&\sum_{j}{\rm Pr}(x_f=e_i,y_f=e_j')
=\sum_{k}\lambda_k|\alpha_{ki}|^2,
\nonumber\\
{\rm Pr}[x_f=j]&=&\sum_{i}{\rm Pr}(x_f=e_i,y_f=e_j')
=\sum_{k}\lambda_k|\beta_{ki}|^2
\label{A59}
\end{eqnarray}
where ${\rm Pr}[x_f=e_i]$ defines a marginal distribution of random variable $x_f$, and ${\rm Pr}[y_f=e_j']$ defines a marginal distribution of random variable $y_f$. It follows that
\begin{eqnarray}
H(x_f,y_f)
&=&-\sum_{i,j}|\gamma_{ij}|^2\log |\gamma_{ij}|^2
\nonumber\\
&\geq& H(x_f)
\label{A60}
\\
&=&-\sum_{i}\sum_{k}\lambda_k|\alpha_{ki}|^2\log \sum_{k}\lambda_k|\alpha_{ki}|^2
\nonumber
\\
&\geq&-\sum_{i}\sum_{k}|\alpha_{ki}|^2 \lambda_k\log \lambda_k
\label{A61}
\\
&=&-\sum_{k}\lambda_k\log \lambda_k
\label{A62}
\\
&=&H(x_0,y_0)
\label{A63}
\end{eqnarray}
Inequality (\ref{A60}) follows from Shannon entropy inequality \cite{Shannon}: $H(x_f,y_f)=H(x_f)+H(y_f|x_f)\geq H(x_f)$ for random variables $x_f,y_f$ on the countable space $\ell^1$. Inequality (\ref{A61}) follows from the concavity of function $f(x)=-x\log{}x$, i.e, $f(\sum_k\gamma_kp_k)\geq \sum_k\gamma_k f(p_k)$ with $\sum_i\gamma_i=1$ and $\gamma_k\geq0$ from Theorem S1. Here, we use $\gamma_k=|\alpha_{ki}|^2$. From the orthogonality of isometric operator ${\cal U}$, we have $\sum_k|\alpha_{ki}|^2=1$ for any $i$, i.e., $\{|\alpha_{ki}|^2, \forall k\}$ is a distribution. Eq.(\ref{A62}) is from the orthogonality of isometric operator ${\cal U}$, i.e., $\sum_i|\alpha_{ki}|^2=1$ for any $k$. Eq.(\ref{A63}) is from Eq.(\ref{A51}). Here, $H(x_f)$ and $H(x_f,y_f)$ can be $+\infty$ for some isometric operators.

Note that Eq.(\ref{A63}) holds for any isometric operators of ${\cal U}:\{|e_i\rangle\}\mapsto \{|f_i\rangle\}$, where $\{|e_i\rangle\}$ and $\{|f_i\rangle\}$ are orthogonal bases. It implies that
\begin{eqnarray}
H(x_0,y_0)&\leq& \min_{{\cal U}, {\cal V}}\{H(x_f,y_f)\}
\nonumber\\
&=&\min_{\hat{J}_a\otimes \hat{J}_b}\{S({\cal D}_{\hat{J}_a\otimes \hat{J}_b}(|F\rangle\langle{}F|))\}
\label{A64}
\end{eqnarray}
where $\hat{J}_a=\{\sum_{k=1}^d\alpha_{ki}|e_i\rangle, \forall i\}$ and $\hat{J}_b=\{\sum_{k=1}^d\beta_{ki}|e_i'\rangle\}$. Moreover, from Eqs.(\ref{A51}) and (\ref{A64}), we have
\begin{eqnarray}
S(\rho)=\min_{J_a\otimes J_b}\{S({\cal D}_{J_a\otimes J_b}(|F\rangle_{ab}\langle{}F|))\}
\label{A65}
\end{eqnarray}
This has proved Eq.(\ref{A45}).

Consider any operator ${\cal U}\otimes \mathbbm{1}$ with ${\cal U}\in \mathbb{SU}(\mathbb{H}_a)$ on the function $|F_0\rangle$ in order to get the function $|F_f\rangle=({\cal U}\otimes \mathbbm{1})|F_0\rangle$. We get that
\begin{eqnarray}
H(y_f)=H(y_0)
\label{A68}
\end{eqnarray}
from Eq.(\ref{A59}). Moreover, $H(y_f|x_f)\geq 0$ for any ${\cal U}\in \mathbb{SU}(\mathbb{H}_a)$. From Eqs.(\ref{A51}), (\ref{A52}) and (\ref{A68}), we get that
\begin{eqnarray}
+\infty>H(y_0)&=&I(x_0;y_0)
\nonumber\\
&\geq & I(x_f;y_0), \forall {\cal U}\in \mathbb{SU}(\mathbb{H}_a)
\label{A69}
\end{eqnarray}

For any isometric operators ${\cal U}\otimes {\cal V}$ with ${\cal U}\in \mathbb{SU}(\mathbb{H}_a)$ and  ${\cal V}\in \mathbb{SU}(\mathbb{H}_b)$, from Data Processing Inequality in Theorem S2, we get
\begin{eqnarray}
I(x_f;y_f)\leq I(x_f;y_0)
\label{A70}
\end{eqnarray}
where $y_f$ is a function of $y_0$ associated with the operator ${\cal V}\in \mathbb{SU}(\mathbb{H}_b)$. Hence, from Eqs.(\ref{A69}) and (\ref{A70}) we have
\begin{eqnarray}
I(x_0;y_0)&=&H(x_0)
\nonumber\\
&\geq& I(x_f;y_f)
\label{A71}
\end{eqnarray}
From Eqs.(\ref{A45}) and (\ref{A71}), we get
\begin{eqnarray}
S(\rho)&=&I(x_0;y_0)
\nonumber\\
&=&\max_{J_a,J_b}\{I(x;y)\}
\label{A72}
\end{eqnarray}
where $J_a=\{\sum_{k}\alpha_{ki}|i\rangle, \forall i\}$ from Eq.(\ref{A55}) and $J_b=\{\sum_{k}\beta_{ki}|i\rangle,\forall i\}$ from Eq.(\ref{A56}). This has proved Eq.(\ref{A46}).

\section{Proof of quantum Renyi entropy}

In this section, we prove that the entropy principle of Renyi entropy. For a given distribution $\{\textrm{Pr}[x=i]=p_i\}_{i=1}^d$ of the random variable $X$, classical Renyi entropy \cite{Renyi} is defined by
\begin{eqnarray}
H_\alpha(X)=\frac{1}{1-\alpha}\log\sum_{i=1}^dp_i^\alpha
\label{B1}
\end{eqnarray}
with $\alpha\geq 0$. This entropy is a generalization of Shannon entropy and satisfies that $\lim_{\alpha\to 1}H_\alpha(X)=H(X)$. For other cases, it reduces to max-entropy for $\alpha=0$, min-entropy for $\alpha=\infty$ and collision entropy for $\alpha=2$. For joint distribution $\{P_{X,Y}(i,j)\}$, Renyi joint entropy satisfies the chain rule:
\begin{eqnarray}
H_\alpha(X,Y)&=&H_\alpha(X)+H_\alpha(Y|X)
\nonumber\\
&=&H_\alpha(Y)+H_\alpha(X|Y)
\label{B2}
\end{eqnarray}
where $H_\alpha(Y|X)$ and $H_\alpha(X|Y)$ denote the Renyi conditional entropies which are defined by
\begin{eqnarray}
H_\alpha(X|Y)=\frac{1}{1-\alpha}\log\frac{\sum_{i,j=1}^dP_{X,Y}(i,j)^\alpha}
{\sum_iP_{Y}(i)^\alpha}
\nonumber\\
H_\alpha(Y|X)=\frac{1}{1-\alpha}\log\frac{\sum_{i,j=1}^dP_{X,Y}(i,j)^\alpha}
{\sum_{i=1}^dP_{X}(i)^\alpha}
\label{B3}
\end{eqnarray}
With the Renyi conditional entropy, Renyi mutual information $I_\alpha(X;Y)$ is defined as
\begin{eqnarray}
I_\alpha(X;Y)
&=&H_\alpha(X)-H_\alpha(X|Y)
\nonumber\\
&=&H_\alpha(Y)-H_\alpha(Y|X)
\nonumber\\
&=&H_\alpha(X)+H_\alpha(Y)-H_\alpha(X,Y)
\label{B4}
\end{eqnarray}

Similar to discussions in Appendix A, consider the separable Hilbert space $\mathbb{H}_a$. For a given system with unknown operator $\rho$ on $\mathbb{H}_a$, where $\{|f_i\rangle\}$ are orthogonal basis on $\mathbb{H}_a$. Define quantum Renyi entropy of the operator $\rho$ as
\begin{eqnarray}
S_\alpha(\rho)
&=&\frac{1}{1-\alpha}\log{\rm tr}\rho^\alpha
\nonumber\\
&=&\frac{1}{1-\alpha}\log\sum_i\lambda_i^\alpha
\label{B28}
\end{eqnarray}
from the spectra decomposition of $\rho=\sum_i\lambda_i|f_i\rangle\langle{}f_i|$. Here,
$S_\alpha(\rho)$ can be $+\infty$ for some Hilbert-Schmidt operator $\rho$. In what follows, we assume that $S_\alpha(\rho)<+\infty$.

Define Renyi entropy of the Hilbert-Schmidt operator ${\cal D}_J(\rho)$ in Eq.(\ref{A35}) as
\begin{eqnarray}
S_\alpha({\cal D}_J(\rho))&=&\frac{1}{1-\alpha}\log{\rm tr}{\cal D}_J(\rho)^\alpha
\nonumber\\
&=&\frac{1}{1-\alpha}\log{}\sum_iq_i^\alpha
\label{B29}
\end{eqnarray}
Our goal in this section is to prove Renyi entropy principle.

Now, consider an operator in Eq.(\ref{A39}) on the separable Hilbert space $\mathbb{H}_a\otimes \mathbb{H}_b$, where $\{|f_{ij}\rangle\}$ are orthogonal basis of $\mathbb{H}_a\otimes \mathbb{H}_b$. Define its partial trace operators $\rho_a$ and $\rho_b$ in Eq.(\ref{A40}). With Eqs.(\ref{A41}) and (\ref{A42}), we can define the mutual information in Eq.(\ref{A43}) under the Renyi entropy in Eq.(\ref{B28}). Define its conditional entropies $S_\alpha(\rho_{a|b})$ and $S_\alpha(\rho_{b|a})$ as
\begin{eqnarray}
&&S_\alpha(\rho_{a|b})=S_\alpha(\rho_{ab})-S_\alpha(\rho_{b})
\label{B30}
\\
&&S_\alpha(\rho_{b|a})=S_\alpha(\rho_{ab})-S_\alpha(\rho_{a})
\label{B31}
\end{eqnarray}
if $\max\{S_\alpha(\rho_{ab}),S_\alpha(\rho_{b}),S(\rho_a)\}<\infty$. Here, we do not define the conditional trace-class operator from the Bayesian rule for avoiding $\infty$ in integral. By using these entropies, we can define the mutual information as
\begin{eqnarray}
I_\alpha(\rho_{ab})
&=&S_\alpha(\rho_{a})-S_\alpha(\rho_{a|b})
\nonumber\\
&=&S_\alpha(\rho_{b})-S_\alpha(\rho_{b|a})
\nonumber\\
&=&S_\alpha(\rho_{a})+S_\alpha(\rho_{b})-S_\alpha(\rho_{ab})
\label{B32}
\end{eqnarray}
if $\max\{S_\alpha(\rho_{a}),S_\alpha(\rho_{b}),S_\alpha(\rho_{ab})\}<+\infty$.

{\bf Theorem S4}(Renyi entropy principle). \textit{For an unknown trace-class operator $\rho$ on separable Hilbert space $\mathbb{H}_a$, consider any one purification $|F\rangle_{ab}$ on separable Hilbert space $\mathbb{H}_a\otimes \mathbb{H}_b$. Then the following statements hold:
\begin{eqnarray}
S_\alpha(\rho)&=&\min_{J}\{S_\alpha({\cal D}_J(\rho))\}
\label{B33}
\nonumber\\
&=&\min_{J_a,J_b}\{S_\alpha({\cal D}_{J_1\otimes J_2}(|F\rangle\langle{}F|))\}
\label{B34}
\nonumber\\
&=&\max_{J_a,J_b}\{I_\alpha({\cal D}_{J_1\otimes J_2}(|F\rangle\langle{}F|))\}
\label{B35}
\end{eqnarray}
where $S_\alpha({\cal D}_{J_1\otimes J_2}(|F\rangle\langle{}F|))$ and $I_\alpha({\cal D}_{J_1\otimes J_2}(|F\rangle\langle{}F|))$ are defined according the joint observed state ${\cal D}_{J_1\otimes J_2}(|F\rangle\langle{}F|)$.
}

{\bf Proof}. The proof is similar to its in Appendix A. Consider an operator $\rho=\sum_{i}\lambda_i|f_i\rangle\langle{}f_i|$ on separable Hilbert space $\mathbb{H}_a$, where $\{|f_i\rangle\}$ are orthogonal functions. Let $|F\rangle=\sum_i\sqrt{\lambda_i}|f_ig_i\rangle$ be a purification of $\rho$ on $\mathbb{H}_a\otimes \mathbb{H}_b$, where $\{|g_i\rangle\}$ are orthogonal functions of $\mathbb{H}_b$. Similar to Eqs.(\ref{A47})-(\ref{A52}), let $\hat{J}_a=\{|f_i\rangle\}$, $\hat{J}_b=\{|g_i\rangle\}$, $J_a=\{|e_i\rangle\}$ and $J_b=\{|e_i'\rangle\}$. From Eqs.(\ref{A48})-(\ref{A52}), we get
\begin{eqnarray}
S_\alpha({\cal D}_{\hat{J}_a\otimes \hat{J}_b}(|F\rangle\langle{}F|))
&=&S_\alpha({\cal D}_{J_a\otimes J_b}(|F_0\rangle\langle F_0|))
\nonumber\\
&=&H_\alpha(x_0,y_0)
\nonumber\\
&=&\frac{1}{1-\alpha}\log\sum_{i}\lambda_i^\alpha
\nonumber\\
&=&H_\alpha(x_0)
\nonumber\\
&=&S_\alpha(\rho)
\label{B36}
\end{eqnarray}
where $H_\alpha(x_0)$ denotes the Renyi entropy and $H_\alpha(x_0,y_0)$ denotes the Renyi joint entropy associated with the distribution in Eq.(\ref{A51}), and $S_\alpha(\rho)$ denotes the quantum Renyi entropy of the operator $\rho$.

Our goal in what follows is to prove that
\begin{eqnarray}
S_\alpha({\cal D}_{J_a\otimes J_b}(|F_0\rangle\langle{}F_0|))=\min_{\hat{J}_a\otimes \hat{J}_b}\{S_\alpha({\cal D}_{\hat{J}_a\otimes \hat{J}_b}(|F\rangle\langle{}F|))\}
\nonumber
\\
\label{B37}
\end{eqnarray}
where $\hat{J}_a$ and $\hat{J}_b$ denote any orthogonal bases on $\mathbb{H}_a$ and $\mathbb{H}_b$, respectively.

From the discussions in Eqs.(\ref{A55}) and (\ref{A59}), we get that
\begin{eqnarray}
H_\alpha(x_f,y_f)
&=&\frac{1}{1-\alpha}\log\sum_{i,j}|\gamma_{ij}|^{2\alpha}
\nonumber\\
&\geq& H_\alpha(x_f)
\label{B38}
\\
&=&\frac{1}{1-\alpha}\log\sum_{i}(\sum_{k}p_k|\alpha_{ki}|^2)^\alpha
\nonumber\\
&\geq & \frac{1}{1-\alpha}\log\sum_{i}|\alpha_{ki}|^2\sum_{k}p_k^\alpha
\label{B39}
\\
&=&\frac{1}{1-\alpha} \log \sum_{k}p_k^\alpha
\label{B40}
\\
&=&H_\alpha(x_0,y_0)
\label{B41}
\end{eqnarray}
Inequality (\ref{B38}) follows from the Renyi entropy inequality: $H_\alpha(x_f,y_f)=H_\alpha(x_f)+H(y_f|y_f)\geq H_\alpha(x_f)$ derived from Eqs.(\ref{B3}) and (\ref{B4}) \cite{Renyi}. Inequality (\ref{B39}) is proved by two facts: (1) $f(x)=\frac{1}{1-\alpha}\log x$ is a decreasing function of $x\in (0,1]$ and $g(p)=p^\alpha$ is a concave function for $\alpha\geq 1$,  i.e, $g(\sum_k\lambda_kp_k)\geq \sum_k\lambda_k g(p_k)$ with $\sum_i\lambda_i=1$ and $\lambda_k\geq0$ from Theorem S1; (2) $f(x)=\frac{1}{1-\alpha}\log x$ is an increasing function of $x\in (0,1]$ and $g(p)=p^\alpha$ is a convex function for $\alpha<1$, i.e, $g(\sum_k\lambda_kp_k)\leq \sum_k\lambda_k g(p_k)$ with $\sum_i\lambda_i=1$ and $\lambda_k\geq0$ from Theorem S1. Here, we use $\lambda_k=|\alpha_{ki}|^2$ because $\{|\alpha_{ki}|^2, \forall k\}$ is a distribution for each $i$ from the orthogonality of ${\cal U}$. Eq.(\ref{B40}) is from the orthogonality of ${\cal U}$, i.e., $\sum_i|\alpha_{ki}|^2=1$ for any $k$. Eq.(\ref{B41}) is from Eq.(\ref{B36}).

Note that Eq.(\ref{B41}) holds for any operator of ${\cal U}:\{|e_i\rangle\}\mapsto \{|f_i\rangle\}$, where $\{|e_i\rangle\}$ and $\{|f_i\rangle\}$ are orthogonal bases. It implies that
\begin{eqnarray}
H_\alpha(x_0,y_0)&\leq & \min_{{\cal U}, {\cal V}}\{H_\alpha(x_f,y_f)\}
\nonumber\\
&=&\min_{\hat{J}_a\otimes \hat{J}_b}\{S_\alpha({\cal D}_{\hat{J}_a\otimes \hat{J}_b}(|F\rangle\langle{}F|))\}
\label{B42}
\end{eqnarray}
where $\hat{J}_a=\{\sum_{k=1}^d\alpha_{ki}|e_i\rangle, \forall i\}$ and $\hat{J}_b=\{\sum_{k=1}^d\beta_{ki}|e_i'\rangle\}$. Moreover, from Eqs.(\ref{A51}) and (\ref{B42}), we have
\begin{eqnarray}
S_\alpha(\rho)=\min_{J_a\otimes J_b}\{S_\alpha({\cal D}_{J_a\otimes J_b}(|F\rangle_{ab}\langle{}F|))\}
\label{B43}
\end{eqnarray}
This has proved the Eq.(\ref{B34}).

Note that from Eqs.(\ref{A51}), (\ref{B38}) and (\ref{B43}), we get
\begin{eqnarray}
S_\alpha(\rho)
&\leq& \min_{{\cal U}}\{H_\alpha(x_f)\}
\nonumber\\
&=&\min_{J}\{S_\alpha({\cal D}_{J}(\rho))\}
\label{B44}
\end{eqnarray}
From Eqs.(\ref{B43}) and (\ref{B44}), it follows that
\begin{eqnarray}
S_\alpha(\rho)=\min_{J}\{S_\alpha({\cal D}_{J}(\rho))\}
\label{B45}
\end{eqnarray}
This has proved Eq.(\ref{B33}).

Consider any operator ${\cal U}\otimes \mathbbm{1}$ with ${\cal U}\in \mathbb{SU}(\mathbb{H}_a)$ on the function $|F_0\rangle$ in order to get the function $|F_f\rangle=({\cal U}\otimes \mathbbm{1})|F_0\rangle$. We get that
\begin{eqnarray}
H_\alpha(y_f)=H_\alpha(y_0)
\label{B46}
\end{eqnarray}
from Eq.(\ref{A59}). Moreover, $H_\alpha(y_f|x_f)\geq 0$ for any ${\cal U}\in \mathbb{SU}(\mathbb{H}_a)$. From Eqs.(\ref{A51}), (\ref{A52}) and (\ref{B46}), we get that
\begin{eqnarray}
+\infty> H_\alpha(x_0)
&=&H_\alpha(y_0)
\nonumber\\
&=&I_\alpha(x_0;y_0)
\nonumber\\
&\geq & I_\alpha(x_f;y_0), \forall {\cal U}\in \mathbb{SU}(\mathbb{H}_a)
\label{B47}
\end{eqnarray}

For any operators ${\cal U}\otimes {\cal V}$ with ${\cal U}\in \mathbb{SU}(\mathbb{H}_a)$ and  ${\cal V}\in \mathbb{SU}(\mathbb{H}_b)$, from Data Processing Inequality in Theorem S2, we get
\begin{eqnarray}
I_\alpha(x_f;y_f)\leq I_\alpha(x_f;y_0)
\label{B48}
\end{eqnarray}
where $y_f$ is a function of $y_0$ associated with the operator ${\cal V}\in \mathbb{SU}(\mathbb{H}_b)$. Hence, from Eqs.(\ref{B47}) and (\ref{B48}) we have
\begin{eqnarray}
I_\alpha(x_0;y_0)\geq& I_\alpha(x_f;y_f)
\label{B49}
\end{eqnarray}
From Eqs.(\ref{A45}) and (\ref{B49}), we get
\begin{eqnarray}
S_\alpha(\rho)
&=&I_\alpha(x_0;y_0)
\nonumber\\
&=&\max_{J_a,J_b}\{I_\alpha(x;y)\}
\label{B50}
\end{eqnarray}
where $J_a=\{\sum_{k}\alpha_{ki}|i\rangle, \forall i\}$ from Eq.(\ref{A55}) and $J_b=\{\sum_{k}\beta_{ki}|i\rangle,\forall i\}$ from Eq.(\ref{A56}). This has proved Eq.(\ref{B35}).

\section{Proof of quantum Tsallis entropy}

In this section, we prove the entropy principle for Tsallis entropy. For a given distribution $\{p_i\}_{i=1}^d$ associated with the random variable $X$, the Tsallis entropy \cite{Tsallis} is defined as
\begin{eqnarray}
T_q(X)=\frac{1}{1-q}(\sum_{i=1}^dp_i^q-1)
\label{C1}
\end{eqnarray}
which tends to Shannon entropy for $q\to 1$, i.e., $\lim_{q\to 1}T_q(X)=H(X)$. Tsallis entropy has its own applications going beyond Shannon entropy \cite{Shannon}. Note that
\begin{eqnarray}
T_q(X)=\frac{1}{1-q}((1-q)\exp(H_q(X))-1)
\label{C2}
\end{eqnarray}
where $H_q(X)$ is Tsallis entropy defined in Eq.(\ref{C1}). Hence, we can define conditional Tsallis entropy as
\begin{eqnarray}
T_q(X|Y)&=&\frac{1}{1-q}((1-q)\exp(H_q(X|Y))-1)
\nonumber\\
&=&\frac{1}{1-q}(\frac{\sum_{i,j=1}^d{\rm Pr}(i,j)^q}{\sum_{i=1}^dp_i^q})
\label{C3}
\end{eqnarray}
where ${\rm Pr}(i,j)$ denotes the joint distribution of random variables $X$ and $Y$, and $p_i$ denotes the marginal distribution of $X$. This definition consists of previous definition \cite{Abe}.

It follows from Eq.(\ref{C3}) that
\begin{eqnarray}
T_q(X,Y)&=&T_q(X)+T_q(Y|X)
+(1-q)T_q(X)T_q(Y|X)
\label{C4}
\end{eqnarray}
This pseudo-additivity property implies that
\begin{eqnarray}
T_q(X,Y)\geq \max\{T_q(X), T_q(Y)\}
\label{C5}
\end{eqnarray}
Unfortunately, the conditional entropy in Eq.(\ref{C3}) is useless for defining Tsallis mutual entropy. Hence, from Eq.(\ref{C4}) the Tsallis mutual information \cite{Tsallis} is defined as
\begin{eqnarray}
I_q(X;Y)
&=&\frac{1}{1+(1-q)\max\{T_q(X),T_q(Y)\}}
\nonumber
\\
&&\times(T_q(X)+T_q(Y)
+(1-q)T_q(X)T_q(Y)
\nonumber
\\
&&
-T_q(X,Y))
\label{C7}
\end{eqnarray}
for any random variables $X$ and $Y$. It is forward to check that
\begin{eqnarray}
I_q(X;Y)=T_q(X)=T_q(Y)
\label{C8}
\end{eqnarray}
for completely dependent variables $X$ and $Y$; and
\begin{eqnarray}
I_q(X;Y)=0
\label{C9}
\end{eqnarray}
for two independent variables $X$ and $Y$.

Similar to discussions in Appendix A, consider the separable Hilbert space $\mathbb{H}_a$. For the given system with unknown operator $\rho$ on $\mathbb{H}_a$, define quantum Tsallis entropy \cite{Tsallis} of the operator $\rho$ as
\begin{eqnarray}
S_q(\rho)
&=&\frac{1}{1-q}({\rm tr}\rho^q-1)
\nonumber\\
&=&\frac{1}{1-q}(\sum_i\lambda_i^q-1)
\label{C29}
\end{eqnarray}
from the spectra decomposition of $\rho=\sum_i\lambda_i|f_i\rangle\langle{}f_i|$, where $\{|f_i\rangle\}$ are orthogonal basis on $\mathbb{H}_a$. Here, $S_q(\rho)$ can be $+\infty$ for some Hilbert-Schmidt operator $\rho$. In what follows, we assume that $S_q(\rho)<+\infty$.

The Tsallis entropy \cite{Tsallis} of the operator ${\cal D}_J(\rho)$ in Eq.(\ref{A35}) is given by
\begin{eqnarray}
S_q({\cal D}_J(\rho))
&=&\frac{1}{1-q}({\rm tr}{\cal D}_J(\rho)^q-1)
\nonumber\\
&=&\frac{1}{1-q}(\sum_iq_i^q-1)
\label{C30}
\end{eqnarray}

From Eqs.(\ref{A39})-(\ref{A42}), define the conditional entropies $S_q(\rho_{a|b})$ and $S_\alpha(\rho_{b|a})$ as
\begin{eqnarray}
&&S_q(\rho_{a|b})=S_q(\rho_{ab})-S_q(\rho_{b})
\label{C31}
\\
&&S_q(\rho_{b|a})=S_q(\rho_{ab})-S_q(\rho_{a})
\label{C32}
\end{eqnarray}
if $\max\{S_q(\rho_{ab}),S_q(\rho_{b}),S(\rho_a)\}<+\infty$.  By using these entropies, we can define the mutual information as
\begin{eqnarray}
I_q(\rho_{ab})&=&S_q(\rho_{a})-S_q(\rho_{a|b})
\nonumber\\
&=&S_q(\rho_{b})-S_q(\rho_{b|a})
\nonumber\\
&=&S_q(\rho_{a})+S_q(\rho_{b})-S_q(\rho_{ab})
\label{C33}
\end{eqnarray}
if $\max\{S_q(\rho_{a}),S_q(\rho_{b}),S_q(\rho_{ab})\}<+\infty$.

{\bf Theorem S5}(Tsallis entropy principle). \textit{For an unknown trace-class operator $\rho$ on separable Hilbert space $\mathbb{H}_a$, consider  any one purification $|F\rangle_{ab}$ on separable Hilbert space $\mathbb{H}_a\otimes \mathbb{H}_b$. Then the Tsallis entropy satisfy the following relations:
\begin{eqnarray}
S_q(\rho)&=&\min_{J}\{S_q({\cal D}_J(\rho))\}
\label{C34}
\\
&=&\min_{J_a,J_b}\{S_q({\cal D}_{J_1\otimes J_2}(|F\rangle\langle{}F|))\}
\label{C35}
\\
&=&\max_{J_a,J_b}\{I_q({\cal D}_{J_1\otimes J_2}(|F\rangle\langle{}F|))\}
\label{C36}
\end{eqnarray}
where $S_q({\cal D}_{J_1\otimes J_2}(|F\rangle\langle{}F|))$ and $I_q({\cal D}_{J_1\otimes J_2}(|F\rangle\langle{}F|))$ are defined according the joint observed state ${\cal D}_{J_1\otimes J_2}(|F\rangle\langle{}F|)$.
}

{\bf Proof}. The proof is similar to Appendix A. Consider an operator $\rho=\sum_{i}\lambda_i|f_i\rangle\langle{}f_i|$ on separable Hilbert space $\mathbb{H}_a$, where $\{|f_i\rangle\}$ are orthogonal basis. Let $|F\rangle=\sum_i\sqrt{\lambda_i}|f_i\rangle|g_i\rangle$ be a purification of $\rho$ on $\mathbb{H}_a\otimes \mathbb{H}_b$, where $\{|g_i\rangle\}$ are orthogonal basis of $\mathbb{H}_b$. Similar to Eqs.(\ref{A47})-(\ref{A52}), let $\hat{J}_a=\{|f_i\rangle\}$, $\hat{J}_b=\{|g_i\rangle\}$, $J_a=\{|e_i\rangle\}$ and $J_b=\{|e_i'\rangle\}$. From Eqs.(\ref{A48})-(\ref{A52}), we get
\begin{eqnarray}
S_q({\cal D}_{\hat{J}_a\otimes \hat{J}_b}(|F\rangle\langle{}F|))
&=&S_q({\cal D}_{J_a\otimes J_b}(|F_0\rangle\langle F_0|))
\nonumber\\
&=&H_q(x_0,y_0)
\nonumber\\
&=&\frac{1}{1-q}(\sum_{i}\lambda_i^q-1)
\nonumber\\
&=&H_q(x_0)
\nonumber\\
&=&S_q(\rho)
\label{C37}
\end{eqnarray}
where $H_q(x_0)$ denotes the Tsallis entropy and $H_q(x_0,y_0)$ denotes the Tsallis joint entropy associated with the distribution in Eq.(\ref{A51}), and $S_q(\rho)$ denotes the quantum Tsallis entropy of the operator $\rho$.

Our goal in what follows is to prove that
\begin{eqnarray}
S_q({\cal D}_{J_a\otimes J_b}(|F_0\rangle\langle{}F_0|))=\min_{\hat{J}_a\otimes \hat{J}_b}\{S_q({\cal D}_{\hat{J}_a\otimes \hat{J}_b}(|F\rangle\langle{}F|))\}
\nonumber
\\
\label{C38}
\end{eqnarray}
where $\hat{J}_a$ and $\hat{J}_b$ denote any orthogonal bases on $\mathbb{H}_a$ and $\mathbb{H}_b$, respectively.

From the discussions in Eqs.(\ref{A55}) and (\ref{A59})  we get that
\begin{eqnarray}
H_q(x_f,y_f)&=&-\sum_{i,j}|\gamma_{ij}|^2\log |\gamma_{ij}|^2
\nonumber\\
&\geq & H_q(x_f)
\label{C39}
\\
&=&-\sum_{i}\sum_{k}p_k|\alpha_{ki}|^2\log \sum_{k}\lambda_k|\alpha_{ki}|^2
\nonumber\\
&\geq & -\sum_{i}\sum_{k}|\alpha_{ki}|^2 \lambda_k\log \lambda_k
\label{C40}
\\
&=&-\sum_{k}\lambda_k\log \lambda_k
\label{C41}
\\
&=&H_q(x_0,y_0)
\label{C42}
\end{eqnarray}
Inequality (\ref{C39}) follows from Tsallis entropy inequality: $H_q(x_f,y_f)=H_q(x_f)+H(y_f|y_f)\geq H_q(x_f)$ derived from Eqs.(\ref{C3}) and (\ref{C4}). Inequality (\ref{C40}) is proved by two facts: (1) $f(x)=\frac{1}{1-q}\log x$ is a decreasing function of $x\in (0,1]$ and $g(p)=p^q$ is a concave function for $q\geq 1$, i.e, $g(\sum_k\lambda_kp_k)\geq \sum_k\lambda_k g(p_k)$ with $\sum_i\lambda_i=1$ and $\lambda_k\geq0$ from Theorem S1; (2) $f(x)=\frac{1}{1-q}\log x$ is an increasing function of $x\in (0,1]$ and $g(p)=p^q$ is a convex function for $q<1$, i.e, $g(\sum_k\lambda_kp_k)\leq \sum_k\lambda_k g(p_k)$ with $\sum_i\lambda_i=1$ and $\lambda_k\geq0$ from Theorem S1. Here, we use $\lambda_k=|\alpha_{ki}|^2$ because $\{|\alpha_{ki}|^2, \forall k\}$ is a distribution for each $i$ from the orthogonality of ${\cal U}$. Eq.(\ref{C41}) is from the orthogonality of ${\cal U}$, i.e., $\sum_i|\alpha_{ki}|^2=1$ for any $k$. Eq.(\ref{C42}) is from Eq.(\ref{C37}). Here, $H_q(x_f,y_f)$ and $H_q(x_f)$ can be $+\infty$ for some quantum measurements.

Note that Eq.(\ref{C42}) holds for any operators of ${\cal U}:\{|e_i\rangle\}\mapsto \{|f_i\rangle\}$, where $\{|e_i\rangle\}$ and $\{|f_i\rangle\}$ are orthogonal bases. It implies that
\begin{eqnarray}
H_q(x_0,y_0)& \leq &
\min_{{\cal U}, {\cal V}}\{H_q(x_f,y_f)\}
\nonumber\\
&=&\min_{\hat{J}_a\otimes \hat{J}_b}\{S_q({\cal D}_{\hat{J}_a\otimes \hat{J}_b}(|F\rangle\langle{}F|))\}
\label{C43}
\end{eqnarray}
where $\hat{J}_a=\{\sum_{k=1}^d\alpha_{ki}|e_i\rangle, \forall i\}$ and $\hat{J}_b=\{\sum_{k=1}^d\beta_{ki}|e_i'\rangle\}$. Moreover, from Eqs.(\ref{A51}) and (\ref{C43}), we have
\begin{eqnarray}
S_q(\rho)=\min_{J_a\otimes J_b}\{S_q({\cal D}_{J_a\otimes J_b}(|F\rangle_{ab}\langle{}F|))\}
\label{C44}
\end{eqnarray}
This has proved the Eq.(\ref{C35}).

Note that from Eqs.(\ref{A51}), (\ref{C39}) and (\ref{C44}), we get
\begin{eqnarray}
S_q(\rho)&\leq& \min_{{\cal U}}\{H_q(x_f)\}
\nonumber\\
&=&\min_{J}\{S_q({\cal D}_{J}(\rho))\}
\label{C45}
\end{eqnarray}
From Eqs.(\ref{C44}) and (\ref{C45}) it follows that
\begin{eqnarray}
S_q(\rho)=\min_{J}\{S_q({\cal D}_{J}(\rho))\}
\label{C46}
\end{eqnarray}
This has proved the Eq.(\ref{C34}).

Consider any operator ${\cal U}\otimes \mathbbm{1}$ with ${\cal U}\in \mathbb{SU}(\mathbb{H}_a)$ on the function $|F_0\rangle$ in order to get the function $|F_f\rangle=({\cal U}\otimes \mathbbm{1})|F_0\rangle$. We get that
\begin{eqnarray}
H_q(y_f)=H_q(y_0)
\label{C47}
\end{eqnarray}
from Eq.(\ref{A59}). Moreover, $H_q(y_f|x_f)\geq 0$ for any operator ${\cal U}\in \mathbb{SU}(\mathbb{H}_a)$. From Eqs.(\ref{A51}), (\ref{A52}) and (\ref{C47}) we get that
\begin{eqnarray}
+\infty> H_q(x_0)
&=&H_q(y_0)
\nonumber\\
&=&I_q(x_0;y_0)
\nonumber\\
&\geq & I_q(x_f;y_0)
\label{C48}
\end{eqnarray}
for any operator ${\cal U}\in \mathbb{SU}(\mathbb{H}_a)$.

For any operators ${\cal U}\otimes {\cal V}$ with ${\cal U}\in \mathbb{SU}(\mathbb{H}_a)$ and  ${\cal V}\in \mathbb{SU}(\mathbb{H}_b)$, from Data Processing Inequality in Theorem S2, we get
\begin{eqnarray}
I_q(x_f;y_f)\leq I_q(x_f;y_0)
\label{C49}
\end{eqnarray}
where $y_f$ is a function of $y_0$ associated with the operator ${\cal V}\in \mathbb{SU}(\mathbb{H}_b)$. Hence, from Eqs.(\ref{C48}) and (\ref{C49}) we have
\begin{eqnarray}
I_q(x_0;y_0)\geq I_q(x_f;y_f)
\label{C50}
\end{eqnarray}
From Eqs.(\ref{A45}) and (\ref{C50}) we get
\begin{eqnarray}
S_q(\rho)&=&I_q(x_0;y_0)
\nonumber\\
&=&\max_{J_a,J_b}\{I_q(x;y)\}
\label{C51}
\end{eqnarray}
where $J_a=\{\sum_{k}\alpha_{ki}|i\rangle, \forall i\}$ from Eq.(\ref{A55}) and $J_b=\{\sum_{k}\beta_{ki}|i\rangle,\forall i\}$ from Eq.(\ref{A56}). This has proved the Eq.(\ref{C36}).

\section{Proof of general quantum entropy}

In this section, we prove general quantum entropy principles. Define a $(d-1)$-dimensional simplex $\Delta^{d-1}$ (finite or indefinite $d$) as
\begin{eqnarray}
\Delta^{d-1}=\left\{
(p_1,\cdots, p_d)\in \mathbb{R}^n: p_i\geq0, \sum_{i=1}^dp_i=1
\right\}
\label{D1}
\end{eqnarray}
For a given distribution $\{p_1, \cdots, p_d\}\in \Delta^{d-1}$ associated with the random variable $X$, a useful entropy is nonnegative function defined on all the distribution space. In general, from the symmetry of $p_i$ it may be represented as
\begin{eqnarray}
H_g(X)=F(\sum_{i=1}^dG(p_i))
\label{D2}
\end{eqnarray}
where $G(\cdot)$ is a function of $\{p_i\}$, and $F(\cdot{})$ is another continuous function. From the axiomatic formulations of Shannon entropy \cite{NS,Furuichi,Tsallis2} or nonextensive entropies \cite{Suyari}, $F$ and $G$ should satisfy some axioms as
\begin{itemize}
\item[(AE1)] \textbf{Continuity}: $G$ is continuous in $\Delta^{d-1}$ and $F$ is continuous on $\mathbb{R}$;
\item[(AE2)] \textbf{Concavity}: $G$ is concave, i.e., $G(\sum_{i=1}^dq_i p_i)\leq \sum_{i=1}^dq_iG(p_i)$ for any distribution $\{q_i\}$;
\item[(AE3)] \textbf{Symmetry}: $G(p_1,\cdots,p_n)$ is symmetric function;
\item[(AE4)] \textbf{Nonnegative}: $F$ is nonnegative;
\item[(AE5)] \textbf{Increasing}: $F$ is an increasing function;
\item[(AE6)] \textbf{Generalized additivity}: For $i=1, \cdots, d,j=1, \cdots, s_i, p_{ij}>0$ and $p_i=\sum_{j=1}^{s_i}p_{ij}$,
\begin{eqnarray}
F(G(p_{11}, \cdots, p_{ds_d}))>F(G(p_{1}, \cdots, p_{d}))
\label{D3}
\end{eqnarray}
\item[(AE7)] \textbf{Expandability}: $F(G(\cdot))$ satisfies the expandability of  $F(G(p_{1}, \cdots, p_{d},0))=F(G(p_{1}, \cdots, p_{d}))$ for any distribution $\{p_{1}, \cdots, p_{d}\}$.
\end{itemize}

Different from previous axiomatic formulations \cite{Tsallis2,Suyari}, the maximality axiom is replaced by the concavity axiomatic. Combining the symmetry of $F(G(\cdot{}))$ and the concavity, it is easy to prove the maximality axiom. Another difference is from the generalized additivity axiom in AE5 which is weaker than previous definition \cite{Suyari}. Since our goal is to present unified axioms for general entropies, we do not require the uniqueness of entropies. From the generalized additivity in Eq.(\ref{D3}), we have
\begin{eqnarray}
H_g(X,Y)\geq \max\{H_g(X),H_g(Y)\}
\label{D4}
\end{eqnarray}
where $H_g(X,Y)$ denotes the entropy defined on joint probability distribution $\{Pr_{XY}(i,j)\}\in \Delta^{d-1}\times \Delta^{d-1}$, $H_g(X)$ and $H_g(Y)$ are defined on marginal distributions, i.e., $p_i(X=i)=\sum_{j}Pr_{XY}(i,j)$ and $p_j(Y=j)=\sum_{i}Pr_{XY}(i,j)$ for all $i,j$. Here, we do not need the additivity of $H_g(X,Y)=H_g(X)+H_g(Y)$ for independent $X$ and $Y$, or the subadditivity of $H_g(X,Y)\leq H_g(X)+H_g(Y)$. The AE7 is used to define quantum generalized entropy.

For our goal in this paper, we need another two measures defined by using $H_g(X)$ in Eq.(\ref{E2}). One is conditional entropy $H_g(X|Y)$ or $H_g(X|Y)$, which is continuous function on two random variables $X$ and $Y$. The conditional entropy is then used to define the mutual information $I_g(X;Y)$ or $I_g(Y;X)$ of two random variables $X,Y$. Generally, it should satisfy the following mutual information axioms:
\begin{itemize}
\item[(AI1)] \textbf{Symmetry}: $I_g(\cdot{};\cdot{})$ is symmetric, i.e.,
$I_g(X;Y)=I_g(Y;X)$ for any two random variables $X$ and $Y$;
\item[(AI2)] \textbf{Consistency}: $I_g(Y;X)\leq \min\{H_g(X),H_g(Y)\}$, and   $I_g(X;Y)=H_g(X)=H_g(Y)$ for completely dependent random variables $X$ and $Y$;
\item[(AI3)]\textbf{Independence}: $I_g(X;Y)=0$ for two independent random variables $X$ and $Y$;
\end{itemize}
Here, we do not require $I_g(X;Y)$ satisfy any additivity of $I_g(X;Y)=H_g(X)-H_g(X|Y)$, $I_g(X;Y)=H_g(Y)-H_g(Y|X)$, or  $I_g(X;Y)=H_g(X)+H_g(Y)-H_g(X,Y)$.

So far, all the well-known entropies \cite{Shannon,Renyi,Tsallis} satisfy the present entropy axioms AE1-AE7. Moreover, one can also define proper mutual information \cite{Shannon,Renyi,Tsallis} which satisfy the axioms AI1-AI3. Another examples are  Havrda-Charvat entropy \cite{HC}, Daroczi entropy \cite{Daroczi}, and Sharma-Mittal information \cite{SM}.

We need the following theorem for defining generalized quantum entropy of Hilbert-Schmidt operator on separable Hilbert space.

\textbf{Theorem S6} \cite{Davis}. \textit{The convex spectral functions are exactly the
symmetric convex functions of the eigenvalues.}

Consider a general Hilbert-Schmidt operator $\rho_{a}=\sum_{i}p_i|\phi_i\rangle\langle \phi_i|$ on separable Hilbert space $\mathbb{H}_a$. From Theorem S6, we can define the generalized quantum entropy of $\rho_{a}$  as
\begin{eqnarray}
S_g(\rho_{a})&=&F({\rm tr}(G(\rho)))
\nonumber\\
&=&F(\sum_{i}G(p_i))
\nonumber\\
&=&H_g(X)
\label{D5}
\end{eqnarray}
where $H_g(X)$ denotes the general entropy of the random variable $X$ defined in Eq.(\ref{D2}). Here, $S_g(\rho)$ can be $+\infty$ for some  Hilbert-Schmidt operator $\rho_a$. In what follows, we assume that $S_g(\rho)<+\infty$.

Define the general entropy of the density operator ${\cal D}_J(\rho)$ in Eq.(\ref{A35}) as
\begin{eqnarray}
S_g({\cal D}_J(\rho))=F(\sum_{i}G(\lambda_i))
\label{D26}
\end{eqnarray}

From Eqs.(\ref{A39})-(\ref{A42}), define the conditional entropies $S_g(\rho_{a|b})$ and $S_\alpha(\rho_{b|a})$ as
\begin{eqnarray}
&S_g(\rho_{a|b})=S_g(\rho_{ab})-S_g(\rho_{b})
\label{D27}
\\
&S_g(\rho_{b|a})=S_g(\rho_{ab})-S_g(\rho_{a})
\label{D28}
\end{eqnarray}
if $\max\{S_g(\rho_{ab}),S_g(\rho_{b}),S_g(\rho_a)\}<+\infty$.  By using these entropies, we can define the mutual information $I_g(\rho_{ab})$ if $\max\{S_g(\rho_{a}),S_g(\rho_{b}),S_g(\rho_{ab})\}<+\infty$.

{\bf Theorem S7} (General entropy principle). \textit{For an unknown trace-class operator $\rho$ with $S_g(\rho_{a})<+\infty$ on separable Hilbert space $\mathbb{H}_a$, consider any one purification $|F\rangle_{ab}$ on separable Hilbert space $\mathbb{H}_a\otimes \mathbb{H}_b$. Then the general entropy satisfies the following relations:
\begin{eqnarray}
S_g(\rho)&=&\min_{J}\{S_g({\cal D}_J(\rho))\}
\label{D29}
\\
&=&\min_{J_a,J_b}\{S_g({\cal D}_{J_1\otimes J_2}(|F\rangle\langle{}F|))\}
\label{D30}
\\
&=&\max_{J_a,J_b}\{I_g({\cal D}_{J_1\otimes J_2}(|F\rangle\langle{}F|))\}
\label{D31}
\end{eqnarray}
where $S_g({\cal D}_{J_1\otimes J_2}(|F\rangle\langle{}F|))$ and $I_g({\cal D}_{J_1\otimes J_2}(|F\rangle\langle{}F|))$ are defined according the joint observed state ${\cal D}_{J_1\otimes J_2}(|F\rangle\langle{}F|)$.
}

{\bf Proof}. The proof is similar to Appendix A. Consider an operator $\rho=\sum_{i}\lambda_i|f_i\rangle\langle{}f_i|$ on separable Hilbert space $\mathbb{H}_a$, where $\{|f_i\rangle\}$ are orthogonal functions. Let $|F\rangle=\sum_i\sqrt{\lambda_i}|f_i\rangle|g_i\rangle$ be a purification of $\rho$ on $\mathbb{H}_a\otimes \mathbb{H}_b$, where $\{|g_i\rangle\}$ are orthogonal functions of $\mathbb{H}_b$. Similar to Eqs.(\ref{A47})-(\ref{A52}), let $\hat{J}_a=\{|f_i\rangle\}$, $\hat{J}_b=\{|g_i\rangle\}$, $J_a=\{|e_i\rangle\}$ and $J_b=\{|e_i'\rangle\}$. From Eqs.(\ref{A48})-(\ref{A52}), we get
\begin{eqnarray}
S_g({\cal D}_{\hat{J}_a\otimes \hat{J}_b}(|F\rangle\langle{}F|))
&=&S_g({\cal D}_{J_a\otimes J_b}(|F_0\rangle\langle F_0|))
\nonumber
\\
&=&H_g(x_0,y_0)
\nonumber
\\
&=&F(\sum_{i}G(\lambda_i))
\nonumber
\\
&=&H_g(x_0)
\nonumber
\\
&=&S_g(\rho)
\label{D32}
\end{eqnarray}
where $H_g(x_0)$ denotes the general entropy and $H_g(x_0,y_0)$ denotes the general joint entropy associated with the distribution in Eq.(\ref{A51}), and $S_g(\rho)$ denotes the general quantum entropy of the operator $\rho$.

Our goal in what follows is to prove that
\begin{eqnarray}
S_g({\cal D}_{J_a\otimes J_b}(|F_0\rangle\langle{}F_0|))=\min_{\hat{J}_a\otimes \hat{J}_b}\{S_g({\cal D}_{\hat{J}_a\otimes \hat{J}_b}(|F\rangle\langle{}F|))\}
\nonumber
\\
\label{D33}
\end{eqnarray}
where $\hat{J}_a$ and $\hat{J}_b$ denote any orthogonal basis functions on $\mathbb{H}_a$ and $\mathbb{H}_b$, respectively.

From the discussions in Eqs.(\ref{A55}) and (\ref{A59}) we get that
\begin{eqnarray}
H_g(x_f,y_f)&=&F(\sum_{i,j}G(|\gamma_{ij}|^2))
\nonumber
\\
&\geq & H_g(x_f)
\label{D34}
\\
&=&F(\sum_{i}G(\sum_{k}p_k|\alpha_{ki}|^2))
\nonumber\\
&\geq & F(\sum_{i}\sum_k|\alpha_{ki}|^2G(p_k))
\label{D35}
\\
&=&F(\sum_kG(p_k))
\label{D36}
\\
&=&H_g(x_0,y_0)
\label{D37}
\end{eqnarray}
Inequality (\ref{D34}) follows from general entropy inequality: $H_q(x_f,y_f)=H_q(x_f)+H(y_f|y_f)\geq H_q(x_f)$. Inequality (\ref{D35}) is from Axioms AE2 and AE4, and Theorem S1. Here, we use $\lambda_k=|\alpha_{ki}|^2$ because $\{|\alpha_{ki}|^2, \forall k\}$ is a distribution for each $i$ from the orthogonality of ${\cal U}$. Eq.(\ref{D36}) is from the orthogonality of ${\cal U}$, i.e., $\sum_i|\alpha_{ki}|^2=1$ for any $k$. Eq.(\ref{D37}) is from Eq.(\ref{D32}). Here, $H_g(x_f,y_f)$ and $H_g(x_f)$ can be $+\infty$ for some quantum projection measurement.

Note that Eq.(\ref{D37}) holds for any operators of ${\cal U}:\{|e_i\rangle\}\mapsto \{|f_i\rangle\}$, where $\{|e_i\rangle\}$ and $\{|f_i\rangle\}$ are orthogonal bases. It implies that
\begin{eqnarray}
H_g(x_0,y_0)
&\leq& \min_{{\cal U}, {\cal V}}\{H_g(x_f,y_f)\}
\nonumber\\
&=&\min_{\hat{J}_a\otimes \hat{J}_b}\{S_g({\cal D}_{\hat{J}_a\otimes \hat{J}_b}(|F\rangle\langle{}F|))\}
\label{D38}
\end{eqnarray}
where $\hat{J}_a=\{\sum_{k=1}^d\alpha_{ki}|e_i\rangle, \forall i\}$ and $\hat{J}_b=\{\sum_{k=1}^d\beta_{ki}|e_i'\rangle\}$. Moreover, from Eqs.(\ref{A51}) and (\ref{D38}), we have
\begin{eqnarray}
S_g(\rho)=\min_{J_a\otimes J_b}\{S_g({\cal D}_{J_a\otimes J_b}(|F\rangle_{ab}\langle{}F|))\}
\label{D39}
\end{eqnarray}
This has proved Eq.(\ref{D30}).

Note that from Eqs.(\ref{A51}), (\ref{D34}) and (\ref{D39}), we get
\begin{eqnarray}
S_g(\rho)&\leq &\min_{{\cal U}}\{H_q(x_f)\}
\nonumber\\
&=&\min_{J}\{S_g({\cal D}_{J}(\rho))\}
\label{D40}
\end{eqnarray}
From Eqs.(\ref{D39}) and (\ref{D40}) it follows that
\begin{eqnarray}
S_g(\rho)=\min_{J}\{S_g({\cal D}_{J}(\rho))\}
\label{D41}
\end{eqnarray}
This has proved Eq.(\ref{D29}).

Consider any operator ${\cal U}\otimes \mathbbm{1}$ with ${\cal U}\in \mathbb{SU}(\mathbb{H}_a)$ on the state $|F_0\rangle$ in order to get $|F_f\rangle=({\cal U}\otimes \mathbbm{1})|F_0\rangle$. We get that
\begin{eqnarray}
H_g(y_f)=H_g(y_0)
\label{D42}
\end{eqnarray}
from Eq.(\ref{A59}). Moreover, $H_g(y_f|x_f)\geq 0$ for any operator ${\cal U}\in \mathbb{SU}(\mathbb{H}_a)$. From Eqs.(\ref{A51}), (\ref{A52}) and (\ref{D42}), we get that
\begin{eqnarray}
+\infty
> H_g(x_0)
=H_g(y_0)
=I_g(x_0;y_0)
\geq I_g(x_f;y_0)
\label{D43}
\end{eqnarray}
for any operator ${\cal U}\in \mathbb{SU}(\mathbb{H}_a)$.

For any operators ${\cal U}\otimes {\cal V}$ with ${\cal U}\in \mathbb{SU}(\mathbb{H}_a)$ and ${\cal V}\in \mathbb{SU}(\mathbb{H}_b)$, from Theorem S2, we get
\begin{eqnarray}
I_g(x_f;y_f)\leq I_g(x_f;y_0)
\label{D44}
\end{eqnarray}
where $y_f$ is a function of $y_0$ associated with the operator ${\cal V}\in \mathbb{SU}(\mathbb{H}_b)$. Hence, from Eqs.(\ref{D43}) and (\ref{D44}) we have
\begin{eqnarray}
I_g(x_0;y_0)\geq I_g(x_f;y_f)
\label{D45}
\end{eqnarray}
From Eqs.(\ref{A45}) and (\ref{D45}), we get
\begin{eqnarray}
S_g(\rho)&=&I_g(x_0;y_0)
\nonumber\\
&=&\max_{J_a,J_b}\{I_g(x;y)\}
\label{D46}
\end{eqnarray}
where $J_a=\{\sum_{k}\alpha_{ki}|i\rangle, \forall i\}$ from Eq.(\ref{A55}) and $J_b=\{\sum_{k}\beta_{ki}|i\rangle,\forall i\}$ from Eq.(\ref{A56}). This has proved the Eq.(\ref{D31}).

\section{Quantum network states violate Shannon entropy inequality}

In Shannon entropy, the joint entropy of two random variable should no less than any one of them \cite{Shannon}, i.e.,
\begin{eqnarray}
H(X_a,X_b)\geq \max\{H(X_a),H(X_b)\}
\label{E1}
\end{eqnarray}
However, for von Neumann entropy of quantum states it only satisfies Araki-Lieb inequality \cite{AL} of
\begin{eqnarray}
S(\rho_{ab})\geq |S(\rho_a)-S(\rho_b)|
\label{E2}
\end{eqnarray}
It means that there are some state $\rho_{AB}$ satisfying $S(\rho_{ab})<\{S(\rho_a),S(\rho_b)\}$.

Our goal in this section is to prove that for any multipartite quantum network consisting of two entangled pure states its von Neumann entropy violates the inequality (\ref{E1}).

Firstly, we prove that von Neumann entropy of tripartite quantum network satisfies the following inequality
\begin{eqnarray}
S(\rho_{a_{i_1}a_{i_2}})<\{S(\rho_{a_1}),S(\rho_{a_2}), S(\rho_{a_3})\},
\label{E3}
\end{eqnarray}
for some $i_1, i_2\in \{1, 2, 3\}$. In fact, consider a tripartite quantum network consisting of two entangled pure states $|\Phi_{1}\rangle$ and $|\Phi_{2}\rangle$. $|\Phi_{1}\rangle$ and $|\Phi_{2}\rangle$ are two bipartite entangled states. In this case, the total state is given by  $|\Phi_{1}\rangle_{a_1a_{2,1}}|\Phi_{2}\rangle_{a_{2,2}a_3}$, where the subsystems $a_{2,1}$ and $a_{2,2}$ are recombined into one subsystem $a_2$. It is easy to get
\begin{eqnarray}
S(\rho_{a_1a_{2}}) &=& S(\rho_{a_3}),
\label{E4}
\\
S(\rho_{a_2a_{3}}) &=& S(\rho_{a_1}),
\label{E5}
\\
S(\rho_{a_1a_{3}}) &=& S(\rho_{a_{2}})
\nonumber\\
&=& S(\rho_{a_{11}})+S(\rho_{a_{12}})
\label{E6}\\
&=& S(\rho_{a_{1}})+S(\rho_{a_{3}})
\label{E7}
\end{eqnarray}
where Eq.(E6) is followed from the additivity of Shannon entropy for tensor state $\rho_{a_{2}}=\rho_{a_{2,1}a_{2,2}}=\rho_{a_{2,1}}\otimes \rho_{a_{2,2}}$. Eq.(\ref{E7}) is followed from the equalities: $S(\rho_{a_{2,1}})=S(\rho_{a_1})$ for any bipartite entanglement $|\Phi_{1}\rangle$, and $S(\rho_{a_{2,2}})=S(\rho_{a_3})$ for any bipartite entanglement $|\Phi_{2}\rangle$.

From Eqs.(\ref{E4})-(\ref{E6}), we have
\begin{eqnarray}
S(\rho_{a_1a_2}) &<& S(\rho_{a_{2}})
\label{E8}
\\
S(\rho_{a_2a_3}) &<& S(\rho_{a_{2}})
\label{E9}
\end{eqnarray}
which have proved the inequality (\ref{E3}).

Generally, consider any acyclic connected $n$-partite quantum entangled network ${\cal N}_q$ consisting bipartite entangled states $|\Phi_{1}\rangle, \cdots, |\Phi_m\rangle$, $n\geq3$. Assume that the $i$-th node $a_i$ consists of $s_i$ systems from $s_i$ bipartite entangled states $|\Phi_{\ell_1}\rangle, \cdots, |\Phi_{\ell_{s_i}}\rangle$. We can prove that the Neumann entropy satisfies the following inequality
\begin{eqnarray}
S(\rho_{a_{i_1}\cdots{}a_{i_{n-1}}})<\{S(\rho_{a_{i_{1}}}),\cdots, S(\rho_{a_{i_{n-1}}})\}
\label{E10}
\end{eqnarray}
for some $i_1,\cdots, i_{n-1}\in \{1, \cdots, n\}$. In fact, it is easy to prove that
\begin{eqnarray}
S(\rho_{a_{i_1}\cdots{}a_{i_{n-1}}})&=&S(\rho_{a_{i_n}})
\label{E11}
\end{eqnarray}
Note that ${\cal N}_q$ is acyclic and connected. It means that $\sum_{i}s_i=2m$. Hence, there is an integer $j$ such that $j=\max\{s_1, \cdots, s_n\}$. From Eq.(\ref{E11}), we have
\begin{eqnarray}
S(\rho_{a_{i_1}\cdots{}a_{i_{n-1}}}) &<& S(\rho_{a_{j}})
\nonumber\\
&=&\max\{S(\rho_{a_{i_{1}}}),\cdots, S(\rho_{a_{i_{n-1}}})\}
\nonumber\\
\label{E12}
\end{eqnarray}
which violates the inequality (\ref{E1}) for $j\in \{i_1,\cdots{}, i_{n-1}\}$. This has completed the proof.

\section{Quantum maximum-entropy principle}

Consider the maximum-entropy principle with von Neumann entropy \cite{Neumann} as follows:
\begin{eqnarray}
&&\mathrm{argmax}_{\{p_i\}}\,\, S(\rho)=-\sum_ip_i\log{}p_i
\nonumber\\
&&\mathrm{s.t.}, \sum_{i=1}^np_i\alpha_{ij}=q_j, j=1, \cdots, m
\nonumber\\
 &&\qquad \sum_{i=1}^np_i=1,
\nonumber\\
 &&\qquad \alpha_{ij}=\langle \phi_i|\psi_j\rangle^2,  i=1, \cdots, m; j=1, \cdots, n.
 \label{F1}
\end{eqnarray}
By rewriting the first restrictions as $ \sum_{i=1}^np_i\frac{\alpha_{ij}}{q_j}=1$ with , one gets a relaxed optimization as
\begin{eqnarray}
&&\mathrm{argmax}_{\{p_i\}}\,\, S(\rho)=-\sum_ip_i\log{}p_i
\nonumber\\
&&\mathrm{s.t.}, \sum_{i=1}^np_i\alpha_i'
\nonumber\\
 &&\qquad \sum_{i=1}^np_i=1,
\nonumber\\
 &&\qquad \alpha_{ij}=\langle \phi_i|\psi_j\rangle^2,  i=1, \cdots, m; j=1, \cdots, n.
 \label{F2}
\end{eqnarray}
with $\alpha_i'=\sum_j\frac{\alpha_{ij}}{q_j}$ for any $i$. This can be easily resolved by using Lagrangian method \cite{Jaynes} as
\begin{eqnarray}
p_i=\exp(-\gamma_1-\gamma_2\alpha_i')
\label{F3}
\end{eqnarray}
where $\gamma_1$ and $\gamma_2$ are Lagrangian multipliers determined by restrictions in Eq.(\ref{F2}). Moreover, the expect entropy $S(\rho)$ is concentrated by the maximum entropy $S_{\max}$ with $\chi^2$ error distribution \cite{Jaynes2}.

Consider the maximum-entropy principle with Renyi entropy \cite{Renyi} as follows:
\begin{eqnarray}
&&\mathrm{argmax}_{\{p_i\}}\,\, S_\alpha(\rho)=\frac{1}{1-\alpha}\log\sum_ip_i^\alpha
\nonumber\\
&&\mathrm{s.t.}, \sum_{i=1}^np_i\alpha_{ij}=q_j, j=1, \cdots, m
\nonumber\\
 &&\qquad \sum_{i=1}^np_i=1,
\nonumber\\
 &&\qquad \alpha_{ij}=\langle \phi_i|\psi_j\rangle^2,  i=1, \cdots, m; j=1, \cdots, n.
 \label{F4}
\end{eqnarray}
Note that $e^{(1-\alpha)S_R(\rho)}=\sum_ip_i^\alpha$ for any distribution $\{p_i\}$. Moreover,  $\sum_ip_i^\alpha\leq 1$ for $\alpha>1$ and  $\sum_ip_i^\alpha\geq 1$ for $0<\alpha<1$. This optimization problem is equivalent to the following problem
\begin{eqnarray}
&&\mathrm{argmax}_{\{p_i\}}\,\, \sum_ip_i^\alpha
\nonumber\\
&&\mathrm{s.t.}, \sum_{i=1}^np_i\frac{\alpha_{ij}}{q_j}=1, j=1, \cdots, m
\nonumber\\
 &&\qquad \sum_{i=1}^np_i=1,
\nonumber\\
 &&\qquad \alpha_{ij}=\langle \phi_i|\psi_j\rangle^2,  i=1, \cdots, m; j=1, \cdots, n.
 \label{F5}
\end{eqnarray}
for $0<\alpha<1$ and
\begin{eqnarray}
&&\mathrm{argmin}_{\{p_i\}}\,\, \sum_ip_i^\alpha
\nonumber\\
&&\mathrm{s.t.}, \sum_{i=1}^np_i\frac{\alpha_{ij}}{q_j}=1, j=1, \cdots, m
\nonumber\\
 &&\qquad\sum_{i=1}^np_i=1,
\nonumber\\
 &&\qquad\alpha_{ij}=\langle \phi_i|\psi_j\rangle^2,  i=1, \cdots, m; j=1, \cdots, n.
 \label{F6}
\end{eqnarray}
for $\alpha>1$. For the optimization in Eq.(\ref{F5}), we can consider a relaxed problem as
\begin{eqnarray}
&&\mathrm{argmax}_{\{p_i\}}\,\, \sum_ip_i^\alpha
\nonumber\\
&&\mathrm{s.t.}, \sum_{i=1}^np_i\alpha'_{i}=1,
\nonumber\\
 &&\qquad \sum_{i=1}^np_i=1,
\nonumber\\
 &&\qquad \alpha_{ij}=\langle \phi_i|\psi_j\rangle^2,  i=1, \cdots, m; j=1, \cdots, n.
 \label{F7}
\end{eqnarray}
with $\alpha'_{i}=\sum_j\frac{\alpha_{ij}}{q_j}$. Now, we use two Lagrange parameters $\gamma_1, \gamma_2$ and define the Lagrange function as
\begin{eqnarray}
F_\alpha=\hat{S}_\alpha+\gamma_1\sum_ip_i+\gamma_1\gamma_2(1-\alpha)\sum_ip_i\alpha'_{i}
\label{F8}
\end{eqnarray}
From $\frac{\partial F_\alpha}{\partial p_i}=0$ for any $i$, we get that
\begin{eqnarray}
p_i=\frac{(1+\gamma_2(1-\alpha)\alpha_i')^{1/\alpha-1}}{\sum_i(1+\gamma_2(1-\alpha)\alpha_i')^{1/\alpha-1}}
\label{F9}
\end{eqnarray}
This distribution is a discrimination of generalized Pareto distribution which derived from an extremum of continuous entropy by using functional Bregman divergence \cite{Pareto}. The same distribution holds for the optimization in Eq.(\ref{F6}). Interestingly, it follows from $\alpha\to 1$ that
\begin{eqnarray}
p_i=\frac{\exp(-\gamma_2\alpha_i')}{\sum_i\exp(-\gamma_2\alpha_i')}
\label{F10}
\end{eqnarray}
which recovers the von Neumann entropy \cite{Neumann}.

Now, consider the maximum-entropy principle with Tsallis entropy \cite{Tsallis} as follows:
\begin{eqnarray}
&&\mathrm{argmax}_{\{p_i\}}\,\, S_q(\rho)=\frac{1}{1-q}(\sum_ip_i^q-1)
\nonumber\\
&&\mathrm{s.t.}, \sum_{i=1}^np_i\alpha_{ij}=q_j, j=1, \cdots, m
\nonumber\\
&&\qquad\sum_{i}p_i=1,
\nonumber\\
&&\qquad \alpha_{ij}=\langle \phi_i|\psi_j\rangle^2,  i=1, \cdots, m; j=1, \cdots, n.
 \label{F11}
\end{eqnarray}
Note that $e^{(1-q)S_q(\rho)}+1=\sum_ip_i^\alpha$ for any distribution $\{p_i\}$. Hence, this optimization problem is equivalent to Eqs.(\ref{F6}) and (\ref{F7}) with solutions in Eq.(\ref{F11}). This correspondence is ensured by the fact that the Renyi entropy is generated by the Tsallis entropy and the additivity of independent systems \cite{Tsallis}.

\section{Proof of Theorem 2}

Consider an unknown quantum source $\rho^{\otimes n}$. Our goal is to find a reliable quantum encoding for $\rho^{\otimes n}$ when $R>S(\rho^c)$, where $\rho^c={\cal D}_J(\rho)$ for some orthogonal basis $J$. Note that for any dephasing operation \cite{Boes19}, there is a unitary transformation $O=\sum_i|i\rangle\langle i|\otimes V_i$ on the space $\mathbb{H}_a\otimes \mathbb{H}_{b}$ satisfying that
\begin{eqnarray}
&&{\rm tr}_{b}[O(\rho\otimes \mathbf{I}_d)O^\dag]=\hat{\rho}
\nonumber\\
&&{\rm tr}_a[O(\rho\otimes \mathbf{I}_d)O^\dag]=\mathbf{I}_d
\label{G1}
\end{eqnarray}
where $V_i$ are unitary operations satisfying ${\rm tr}[V_iV_j]=d\delta_{ij}$, and $\mathbb{H}_{b}$ is an axillary space. For general source $\rho$, from Eq.(\ref{G1}) we get
\begin{eqnarray}
&&{\rm tr}_{b_1\cdots b_n}[O^{\otimes n}(\rho^{\otimes n}\otimes \mathbf{I}_d^{\otimes n})(O^{\otimes n})^\dag]=(\rho^c)^{\otimes n}
\nonumber\\
&&{\rm tr}_{a_1\cdots a_n}[O^{\otimes n}(\rho^{\otimes n}\otimes \mathbf{I}_d^{\otimes n})(O^{\otimes n})^\dag]=\mathbf{I}_d^{\otimes n}
\label{G2}
\end{eqnarray}

In what follows, we take use of universal quantum information compression \cite{Jozsa} based on classical scheme \cite{Cziszar}. In fact, consider a classical distribution $\{p_i\}$ with Shannon entropy $H(X)$. From Shannon Theorem \cite{Shannon} there is a typical series $T^{(n)}_\epsilon$  with asymptotic equipartition, where $T^{(n)}_\epsilon=\{x_1\cdots x_n\}$ satisfying
\begin{eqnarray}
2^{-n(H(X)+\epsilon)}\leq p(x_1,\cdots, x_n)\leq 2^{-n(H(X)-\epsilon)}
\label{G3}
\end{eqnarray}
for any given $\epsilon>0$. $T^{(n)}_\epsilon$ is further reduced into a subset $C^{(n)}_\epsilon \subseteq T^{(n)}_\epsilon$ \cite{Cziszar} which satisfies all of the properties of $T^{(n)}_\epsilon$ for any probability distribution with Shannon entropy no more than $S$. It meant that
\begin{eqnarray}
&&{\rm Pr}[C^{(n)}_\epsilon]>1-\epsilon, \forall \{p_i\}
\label{G4}
\\
&&|C^{(n)}_\epsilon|=2^{n(S+\delta)}
\label{G5}
\end{eqnarray}

Let $\{\lambda_i\}$ denote the spectra of $\rho$. The spectra of $(\rho^c)^{\otimes n}$ is given by $\{\lambda_{i_1\cdots i_n}= \lambda_{i_1}\cdots \lambda_{i_n}\}$. Let $\Lambda_n$ be the subspace of $\mathbb{H}^{\otimes n}$, i.e.,
\begin{eqnarray}
\Lambda_n={\rm span}\{|\lambda_{i_1\cdots i_n}\rangle| i_1\cdots i_n\in C^{(n)}_\epsilon\}
\label{G6}
\end{eqnarray}
Note that $H(\lambda_i)=S(\rho^c)$. It means that
\begin{eqnarray}
{\rm dim}(\Lambda_n)=2^{n(S+\delta)}
\label{G7}
\end{eqnarray}
 i.e. the typical subspace is $nS(\rho^c)$ qubits. Let $\Pi$ denote the projection onto $\Lambda_n$. From Eq.(\ref{G4}) we get that
\begin{eqnarray}
{\rm tr}[\Pi(\rho^c)^{\otimes n}]>1-\epsilon
\label{G8}
\end{eqnarray}
It means that the projection onto $\Lambda_n$ presents a faithful compression for $(\rho^c)^{\otimes n}$ \cite{Jozsa}. From Eq.(\ref{G2}), $O^{\otimes n}$ is unitary. This implies that the projection onto $\Lambda_n$ gives a faithful compression for $\rho^{\otimes n}$. Moreover, for any source $\rho'$ with $S(\rho')\leq S(\rho)$, from Theorem 1, we have $S(\rho')\leq S(\rho^c)$. Moreover, ${\cal D}_J(\rho')$ and ${\cal D}_J(\rho)$ are commute. If $R>S({\cal D}_J(\rho'))$, the projection onto $\Lambda_n$ is a faithful compression for ${\cal D}_J(\rho')^{\otimes n}$ \cite{Jozsa}. From Eq.(\ref{G2}), it is a faithful compression for $\rho'^{\otimes n}$. This provides a weak universal quantum information compression for unknown sources.

\section{Proof of Theorem 3}

{\bf Proof of Theorem 3}. Consider the specific experimental state $\rho^c$ with the computation basis $J=\{|j\rangle_a\}$. Note that $\rho^c={\cal D}_J(\rho)$. From the dephasing lift lemma \cite{Boes19}, there exists a unitary $U_1=\sum_{j}|j\rangle\langle j|\otimes V_j$ such that
\begin{eqnarray}
&&{\rm tr}_{b}[U_1(\rho\otimes \mathbf{1}_{b})U_1^\dag]={\cal D}_j(\rho)=\rho^c
\label{H1}
\\
&&{\rm tr}_a[U_1(\rho\otimes \mathbf{1}_{b})U_1^\dag]=\mathbf{1}_{b}
\label{H2}
\end{eqnarray}
where $V_j$ satisfies ${\rm tr}[V_iVj]=d\delta_{ij}$, and $d$ denotes the dimension of Hilbert space $\mathbb{H}_a$.

If $S(\rho^c)>S(\rho')$ and ${\rm rank}(\rho')\geq {\rm rank}(\rho^c)$, it follows that  there exists a density matrix $\tau_b$ and another unitary matrix $U_2$ such that \cite{Boes19}:
\begin{eqnarray}
&&{\rm tr}_{b'}[U_2(\rho\otimes \tau_{b'})U_2^\dag]=\rho'
\label{H3}
\\
&&{\cal D}_J[{\rm tr}_a(U_2(\rho\otimes \tau_{b'})U_2^\dag)]=\tau_{b'}
\label{H4}
\end{eqnarray}
Now, define
\begin{eqnarray}
&&\sigma_{bb'}=\mathbf{1}_{b}\otimes \tau_{b'},
\nonumber\\
&&U=(U_2\otimes \mathbbm{1}_{b})(U_1\otimes \mathbbm{1}_{b'})
\label{H5}
\end{eqnarray}
Note that $U_2$ has not changed the system $b'$ while $U_1$ has not changed the system $b$. It means that $U_2$ and ${\rm tr}_{b'}(\cdot)$ ($U_1$ and ${\rm tr}_{b}(\cdot)$) are commute. From Eqs.(\ref{H1}) and (\ref{H2}), it follows that
\begin{eqnarray}
&&{\rm tr}_{bb'}[U(\rho\otimes \sigma_{bb'})U^\dag]
\nonumber
\\
&=&{\rm tr}_{b'}[U_2 ({\rm tr}_b((U_1\otimes \mathbbm{1}_{b'})
(\rho\otimes \sigma_{bb'})(U_1^\dag\otimes \mathbbm{1}_{b'})))U_2^\dag]
\nonumber\\
&=&{\rm tr}_{b'}[U_2 (\rho_a \otimes \tau_{b'})U_2^\dag]
\nonumber\\
&=&\rho'
\label{H6}
\end{eqnarray}
Note that the systems $b$ and $b'$ are uncorrelated after $U_1$ being performed. Moreover, $U_2$ has not affected the systems $b$ and $b'$. It means that $b$ and $b'$ are always uncorrelated. Hence, from Eqs.(\ref{H2}) and (\ref{H4}), it follows that
\begin{eqnarray}
{\cal D}_J[{\rm tr}_{a}(U(\rho\otimes \sigma)U^\dag)]=\sigma
\label{H7}
\end{eqnarray}
This has completed the proof. $\Box$

Theorem 3 also holds for the majorization relation as follows.

\textbf{Theorem S8}. \textit{If $\rho^c\succeq\rho'$ and $\rho^c$ and $\rho'$ have different spectra, there exists a unitary $U$ such that
\begin{eqnarray}
&&{\rm tr}_b[U(\rho\otimes \mathbf{1}_{d^2})U^\dag]=\rho',
\nonumber\\
&&{\rm tr}_a[U(\rho\otimes \mathbf{1}_{d^2})U^\dag]=\mathbf{1}_{d^2}
\label{H8}
\end{eqnarray}
where $\mathbf{1}_{d^2}$ is the maximally mixed state on Hilbert space $\mathbb{H}_b$.}

\textbf{Lemma S1} \cite{Gour}. \textit{Let $\rho$ and $\rho'$ be two density matrices on Hilbert space $\mathbb{H}_a$. If $\rho' \preceq \rho$, there exists a unitary $U$ such that
\begin{eqnarray}
&&{\rm tr}_b[U(\rho\otimes \mathbf{1}_{b})U^\dag]=\rho'
\nonumber\\
&&{\rm tr}_a[U(\rho\otimes \mathbf{1}_{b})U^\dag]=\mathbf{1}_{b}
\label{H9}
\end{eqnarray}
where $\mathbf{1}_{R}$ denotes the maximally mixed state of an axillary system $R$.}

{\bf Proof of Theorem S8}. It only needs to consider the specific state $\rho^c$ with the computation basis $J=\{|j\rangle_A\}$.  Note that $\rho^c={\cal D}_J(\rho)$. From the dephasing lift lemma \cite{Boes19}, there exists a unitary matrix $U_1=\sum_{j}|j\rangle\langle j|\otimes V_j$ such that
\begin{eqnarray}
&&{\rm tr}_{b}[U_1(\rho\otimes \mathbf{1}_{b})U_1^\dag]=D_j(\rho)=\rho^c
\label{H10}\\
&&{\rm tr}_a[U_1(\rho\otimes \mathbf{1}_{b})U_1^\dag]=\mathbf{1}_{b}
\label{H11}
\end{eqnarray}
where $V_j$ satisfies ${\rm tr}[V_iV_j]=d\delta_{ij}$, and $d$ denotes the dimension of Hilbert space $\mathbb{H}_a$.

From Lemma S1, there exists a unitary matrix $U_2$ such that
\begin{eqnarray}
&&{\rm tr}_{b'}[U_2(\rho\otimes \mathbf{1}_{b'})U_2^\dag]=\rho'
\label{H12}\\
&&{\rm tr}_a[U_2(\rho\otimes \mathbf{1}_{b'})U_2^\dag]=\mathbf{1}_{b'}
\label{H13}
\end{eqnarray}
where $b'$ is a $d$-dimensional axillary system.

Now, define
\begin{eqnarray}
&&\mathbf{1}_{d^2}=\mathbf{1}_{b}\otimes \mathbf{1}_{b'},
\nonumber\\
&&U=(U_2\otimes \mathbbm{1}_{b})(U_1\otimes \mathbbm{1}_{b'})
\label{H14}
\end{eqnarray}
Note that $U_2$ has not changed the system $b'$ while $U_1$ has not changed the system $b$. It means that $U_2$ and ${\rm tr}_{b'}[\cdot]$ ($U_1$ and ${\rm tr}_{b}[\cdot]$) are commute. From Eqs.(\ref{H12}) and (\ref{H14}), it follows that
\begin{eqnarray}
&&{\rm tr}_{bb'}[U(\rho\otimes \mathbf{1}_{d^2})U^\dag]
\nonumber
\\
&=&{\rm tr}_{b'}[U_2 ({\rm tr}_b((U_1\otimes \mathbbm{1}_{b'})
(\rho\otimes \sigma)(U_1^\dag\otimes \mathbbm{1}_{b'})))U_2^\dag]
\nonumber\\
&=&{\rm tr}_{b'}[U_2 (\rho_a \otimes \mathbbm{1}_{b'})U_2^\dag]
\nonumber\\
&=&\rho'
\label{H15}
\end{eqnarray}
Note that the systems $b$ and $b'$ are uncorrelated after $U_1$ being performed. Moreover, $U_2$ has not affected the systems $b$ and $b'$. It means that $b$ and $b'$ are always uncorrelated. Hence, from Eqs.(\ref{H13}) and (\ref{H14}), it follows that
\begin{eqnarray}
{\rm tr}_{a}[U(\rho\otimes \mathbf{1}_{d^2})U^\dag]=\mathbf{1}_{d^2}
\label{H16}
\end{eqnarray}
This completes the proof. $\Box$

Similar proof holds for $\rho^c\succeq_T\rho'$ \cite{Gour}.

\section{Approximate transition of infinite dimensional states}

Consider a system $a$ on separable Hilbert space $\mathbb{H}_a$. Our goal in this section is to consider the approximate transition of state $\rho_a\to \rho'_a$ under the local operations.

\textbf{Definition S1} (Approximate transition) \cite{Owari}. A infinite dimensional system $\rho_a$ is $\epsilon$-transition to $\rho'$ by using local operation ${\cal E}$ if $\|{\cal E}(\rho)-\rho'\|_{tr}<\epsilon$, where $\|\cdot\|_{tr}$ denotes the trace norm.

\textbf{Theorem S9}. \textit{Let $\rho_c$ and $\rho'$ be two Hilbert-Schmidt operators on separable Hilbert space $\mathbb{H}_a$. If $\rho^c\succeq\rho'$ and $\rho^c$ and $\rho'$ have different spectra, for any small constant $\epsilon>0$ there exist a unitary $U$ and finite-dimensional state $\mathbf{1}_{R}$ such that
\begin{eqnarray}
&&\|{\rm Tr}_R[U(\rho\otimes \mathbf{1}_{R})U^\dag]-\rho'\|_{tr}<\epsilon
\nonumber\\
&&\|{\rm Tr}_A[U(\rho\otimes \mathbf{1}_{R})U^\dag]-\mathbf{1}_{R}\|_{tr}<\epsilon
\label{I1}
\end{eqnarray}
where $\mathbf{1}_{R}$ denotes the maximally mixed state of a finite-dimensional  axillary system $R$.}

{\bf Proof}. Let $\rho^c$ and $\rho'$ be two Hilbert-Schmidt operators on separable Hilbert space $\mathbb{H}_a$. Suppose that the spectra decomposition of $\rho^c, \rho'$ are given by
\begin{eqnarray}
\rho^c=\sum_{i=1}^\infty p_i|\phi_i\rangle\langle \phi_i|,
\nonumber\\
\rho'=\sum_{i=1}^\infty q_i|\psi_i\rangle\langle \psi_i|,
\label{I2}
\end{eqnarray}
For simplicity, assume that $p_i$ and $q_i$ are decreasing series in term of $i=1, \cdots$. Now, consider the unitary operators $U:|\phi_i\rangle\to |i\rangle$ and $V:|\psi_i\rangle\to |i\rangle$. $\rho^c$ and $\rho'$ can be changed into normal forms as
\begin{eqnarray}
U\rho^cU^\dag=\sum_{i=1}^\infty p_i|i\rangle\langle i|,
\nonumber\\
V\rho'V^\dag=\sum_{i=1}^\infty q_i|i\rangle\langle i|.
\label{I3}
\end{eqnarray}
Note that $\|\rho^c\|_{tr}=\|\rho'\|_{tr}=1$.  So, for a given small constant $\epsilon>0$, there exists an integer $n$ such that
\begin{eqnarray}
\|\rho^c-(\rho^c)^{(n)}\|_{tr}&=&\|U\rho^cU^\dag-\sum_{i=1}^n p_iU^\dag|\phi_i\rangle\langle \phi_i|U^\dag\|_{tr}
\nonumber
\\
&<&\frac{\epsilon}{4}
\label{I4}
\\
\|\rho'-\rho'^{(n)}\|_{tr}
&=&\|V\rho'V^\dag-\sum_{i=1}^n q_iV^\dag|\psi_i\rangle\langle \psi_i|V^\dag\|_{tr}
\nonumber
\\
&<&\frac{\epsilon}{4}
\label{I5}
\end{eqnarray}
where $(\rho^c)^{(n)}$ and $\rho'^{(n)}$ are defined by
\begin{eqnarray}
(\rho^c)^{(n)}=\sum_{i=1}^n p_i|\phi_i\rangle\langle \phi_i|
\nonumber\\
\rho'^{(n)}=\sum_{i=1}^n q_i|\psi_i\rangle\langle \psi_i|
\label{I6}
\end{eqnarray}
Note that $(\rho^c)^{(n)} \succeq \rho'^{(n)}$ from $\rho_c\succeq\rho'$, where we do not require the equality of ${\rm tr}\rho_c^{(n)}={\rm tr} \rho'^{(n)}$. From Theorem S8, there exists a unitary $W$ on $n^3$-dimensional Hilbert space $\mathbb{H}_n\otimes \mathbb{H}_{n^2}$ such that
\begin{eqnarray}
&&{\rm tr}_b[W((\rho^c)^{(n)}\otimes \mathbf{1}_{n^2})W^\dag]=\frac{{\rm tr} (\rho^c)^{(n)}}{{\rm tr} \rho'^{(n)}}\rho'^{(n)},
\label{I7}\\
&&{\rm tr}_a[W((\rho^c)^{(n)}\otimes \mathbf{1}_{n^2})W^\dag]={\rm tr} (\rho^c)^{(n)}\mathbf{1}_{d^2}
\label{I8}
\end{eqnarray}
where $\mathbf{1}_{n^2}$ is the maximally mixed state of rank $n^2$. Now, we can extend $W$ into a unitary operator $\hat{W}=W\oplus \mathbb{I}_r$ on the separable space $\mathbb{H}_a$, where $\mathbbm{1}_r$ denotes the identity operator on the orthogonal complement space of $\mathbb{H}_n\otimes \mathbb{H}_{n^2}$ in $\mathbb{H}_a$. Since the trace norm is invariant under the unitary operator. It follows that
\begin{eqnarray}
&&\|{\rm tr}_b[\hat{W}(\rho^c\otimes \mathbf{1}_{n^2})\hat{W}^\dag]-
{\rm tr}_b[W(\frac{(\rho^c)^{(n)}}{{\rm tr} (\rho^c)^{(n)}}\otimes \mathbf{1}_{n^2})W^\dag]\|_{tr}
\nonumber\\
&=&
\|\hat{W}(\rho^c\otimes \mathbf{1}_{n^2})\hat{W}^\dag-
W(\frac{(\rho^c)^{(n)}}{{\rm tr} (\rho^c)^{(n)}}\otimes \mathbf{1}_{n^2})W^\dag\|_{tr}
\nonumber
\\
&=&
\|\rho_c\otimes \mathbf{1}_{n^2}-
\frac{(\rho^c)^{(n)}}{{\rm tr} (\rho^c)^{(n)}}\otimes \mathbf{1}_{n^2}\|_{tr}
\nonumber
\\
&=&
\|\rho_c-\frac{(\rho^c)^{(n)}}{{\rm tr} (\rho^c)^{(n)}}\|_{tr}
\nonumber\\
&\leq &
\|\rho^c-(\rho^c)^{(n)}\|_{tr}
+\|(\rho^c)^{(n)}-\frac{(\rho^c)^{(n)}}{{\rm tr} (\rho^c)^{(n)}}\|_{tr}
\nonumber\\
&\leq & \frac{\epsilon}{2}
\label{I9}
\end{eqnarray}
from the triangle inequality of trace norm and inequality (\ref{I4}). Moreover, we get
\begin{eqnarray}
\|\frac{\rho'^{(n)}}{{\rm tr}\rho'^{(n)}}-\rho'\|_{tr}
&\leq &
\|\frac{\rho'^{(n)}}{{\rm tr}\rho'^{(n)}}-\rho'^{(n)}\|_{tr}
+\|\rho'^{(n)}-\rho'\|_{tr}
\nonumber
\\
&\leq & \frac{\epsilon}{2}
\label{I10}
\end{eqnarray}
from the triangle inequality of trace norm and inequality (\ref{I5}). From Eqs.(\ref{I4}),(\ref{I6}), (\ref{I9}) and (\ref{I10}) we have
\begin{eqnarray}
&&\|{\rm tr}_b[\hat{W}(\rho^c\otimes \mathbf{1}_{n^2})\hat{W}^\dag]-\rho'\|_{tr}
\nonumber
\\
&\leq &
\|{\rm tr}_b[\hat{W}(\rho^c\otimes \mathbf{1}_{n^2})\hat{W}^\dag]-
{\rm tr}_b[W(\frac{(\rho^c)^{(n)}}{{\rm tr} (\rho^c)^{(n)}}\otimes \mathbf{1}_{n^2})W^\dag]\|_{tr}
\nonumber\\
&&+\|{\rm tr}_b[W(\frac{(\rho^c)^{(n)}}{{\rm tr} (\rho^c)^{(n)}}\otimes \mathbf{1}_{n^2})W^\dag]-\frac{\rho'^{(n)}}{{\rm tr}\rho'^{(n)}}\|_{tr}
\nonumber\\
&&+\|\frac{\rho'^{(n)}}{{\rm tr}\rho'^{(n)}}-\rho'\|_{tr}
\nonumber\\
&<&\epsilon
\label{I11}
\end{eqnarray}
and
\begin{eqnarray}
&&\|{\rm tr}_a[\hat{W}(\rho^c\otimes \mathbf{1}_{n^2})\hat{W}^\dag]-\mathbf{1}_{d^2}\|_{tr}
\nonumber
\\
&\leq &
\|{\rm tr}_a[\hat{W}(\rho^c\otimes \mathbf{1}_{n^2})\hat{W}^\dag]-
{\rm tr}_a[W(\frac{(\rho^c)^{(n)}}{{\rm tr} (\rho^c)^{(n)}}\otimes \mathbf{1}_{n^2})W^\dag]\|_{tr}
\nonumber\\
&&+\|{\rm tr}_a[W(\frac{(\rho^c)^{(n)}}{{\rm tr} (\rho^c)^{(n)}}\otimes \mathbf{1}_{n^2})W^\dag]-\mathbf{1}_{d^2}\|_{tr}
\nonumber\\
&<&\epsilon
\label{I12}
\end{eqnarray}
This has completed the proof. $\Box$.

Different from previous approximate transition \cite{Owari}, the axillary system is finite dimensional. Similar result holds for entropy condition.

\textbf{Theorem S10}. \textit{Let $\rho^c$ and $\rho'$ be two Hilbert-Schmidt operators on separable Hilbert space $\mathbb{H}_a$. If $S(\rho')<S(\rho^c)<\infty$, for any small constant $\epsilon>0$ there exists a unitary $U$ and a finite density matrix $\sigma_b$ on Hilbert space $\mathbb{H}_b$ such that
\begin{eqnarray}
&&\|{\rm tr}_b[U(\rho\otimes \sigma)U^\dag]-\rho'\|_{tr}<\epsilon,
\\
&&\|{\cal D}_J[{\rm tr}_a(U(\rho\otimes \sigma)U^\dag)]-\sigma\|_{tr}<\epsilon
\label{I13}
\end{eqnarray}
where $S(\rho)$ denotes the von Neumann entropy.}

{\bf Proof of Theorem S10}. Let $\rho^c$ and $\rho'$ be two Hilbert-Schmidt operators on separable Hilbert space $\mathbb{H}_a$. From Eq.(\ref{I5}), we have $S(\rho_c^{(n)}) > S(\rho'^{(n)})$ from $S(\rho')<S(\rho^c)<\infty$, where we do not require the equality of ${\rm tr}(\rho^c)^{(n)}$ and ${\rm tr} \rho'^{(n)}$ to be normalized. From Theorem 3, there exist a finite-dimensional density matrix $\sigma_b$ on Hilbert space $\mathbb{H}_b$ and a unitary $W$ on $\mathbb{H}_n\otimes \mathbb{H}_{n^2}$ such that
\begin{eqnarray}
&&{\rm tr}_r[W((\rho^c)^{(n)}\otimes \sigma_b)W^\dag]=\frac{{\rm tr} (\rho^c)^{(n)}}{{\rm tr} \rho'^{(n)}}\rho'^{(n)},
\label{I14}
\\
&&{\cal D}_J[{\rm tr}_a(W(\frac{(\rho^c)^{(n)}}{{\rm tr} (\rho^c)^{(n)}}\otimes \sigma_b)W^\dag)]=\sigma_b
\label{I15}
\end{eqnarray}
Now, we can extend $W$ into a unitary operator $\hat{W}=W\oplus \mathbb{I}_r$ on the separable space $\mathbb{H}_a$, where $\mathbbm{1}_r$ denotes the identity operator on the orthogonal complement space of $\mathbb{H}_n\otimes \mathbb{H}_{n^2}$ in $\mathbb{H}_a$. Similar to the inequality (\ref{I9}), from the inequality (\ref{I4}) and Eq.(\ref{I14}) we get that
\begin{eqnarray}
&&\|{\rm tr}_b[\hat{W}(\rho^c\otimes \sigma_b)\hat{W}^\dag]-
{\rm tr}_b[W(\frac{(\rho^c)^{(n)}}{{\rm tr} (\rho^c)^{(n)}}\otimes \sigma_b)W^\dag]\|_{tr}
\nonumber\\
&=&
\|\rho^c\otimes \sigma_b-
\frac{(\rho^c)^{(n)}}{{\rm tr}(\rho^c)^{(n)}}\otimes \sigma_b\|_{tr}
\nonumber
\\
&\leq &
\|\rho^c-(\rho^c)^{(n)}\|_{tr}
+\|(\rho^c)^{(n)}-\frac{(\rho^c)^{(n)}}{{\rm tr}(\rho^c)^{(n)}}\|_{tr}
\nonumber\\
&\leq & \frac{\epsilon}{2}
\label{I16}
\end{eqnarray}
From Eqs.(\ref{I4}), (\ref{I5}), (\ref{I14})-(\ref{I16}), we have
\begin{eqnarray}
&&\|{\rm tr}_b[\hat{W}(\rho^c\otimes \sigma_b)\hat{W}^\dag]-\rho_b'\|_{tr}
\nonumber\\
&\leq &
\|{\rm tr}_b[\hat{W}(\rho^c\otimes  \sigma_b)\hat{W}^\dag]-
{\rm tr}_2[W(\frac{(\rho^c)^{(n)}}{{\rm tr} (\rho^c)^{(n)}}\otimes \sigma_b)W^\dag]\|_{tr}
\nonumber\\
&&+\|{\rm tr}_b[W(\frac{(\rho^c)^{(n)}}{{\rm tr} (\rho^c)^{(n)}}\otimes \sigma_b)W^\dag]-\frac{\rho'^{(n)}}{{\rm tr}\rho'^{(n)}}\|_{tr}
\nonumber\\
&&
+\|\frac{\rho'^{(n)}}{{\rm tr}\rho'^{(n)}}-\rho'\|_{tr}
\nonumber\\
&<&\epsilon
\label{I17}
\end{eqnarray}
and
\begin{eqnarray}
&&\|{\cal D}_J[{\rm tr}_a(\hat{W}(\rho^c\otimes \sigma)\hat{W}^\dag)]-\sigma\|_{tr}
\nonumber\\
&\leq &
\|{\cal D}_J[{\rm tr}_a(\hat{W}(\rho^c\otimes \sigma)\hat{W}^\dag)]
\nonumber\\
&&-
{\cal D}_J[{\rm tr}_a(W(\frac{(\rho^c)^{(n)}}{{\rm tr}(\rho^c)^{(n)}}\otimes \sigma)W^\dag)]\|_{tr}
\nonumber\\
&&+\|{\cal D}_J[{\rm tr}_a(W(\frac{(\rho^c)^{(n)}}{{\rm tr} (\rho^c)^{(n)}}\otimes \sigma)W^\dag)]-\sigma\|_{tr}
\nonumber\\
&<&\epsilon
\label{I18}
\end{eqnarray}
where the trace norm is invariant for the dephasing operation ${\cal D}_J$ which can regarded as the combination of trace operation and unitary transformation. This completes the proof. $\Box$

\section{Any state conversion}

\textbf{Theorem S11} (Any state conversion). \textit{For any two states $\rho_a, \rho'_a$ on Hilbert space $\mathbb{H}_a$, there exists a unitary operation $U$ dependent on $\rho$ such that
\begin{eqnarray}
&&{\rm Pr}[{\rm tr}_b[U(\rho \otimes \mathbf{I}_b)U^\dag]=\rho']=p,
\nonumber\\
&&{\rm tr}_a[U(\rho\otimes \mathbf{I}_b)U^\dag]=\mathbf{I}
\label{K1}
\end{eqnarray}
if $\rho \succeq p\rho'$ with $p\in (0,1)$, where $\mathbf{I}$ denotes the maximally mixed state on Hilbert space $\mathbb{H}_b$.}

{\bf Proof}. From the locality, the local unitary can be performed after the catalytic operation. Hence, it only needs to consider the diagonal density matrices $\rho$ and $\rho'$. Assume that ${\rm rank}(\rho')=d$, i.e., $\rho'$ has full rank. Otherwise, one define the reduced matrix without zero spectra. Define the density matrix
\begin{eqnarray}
\hat{\rho}=
\left(
\begin{array}{ccc}
p\rho' & 0
\nonumber\\
0 & \mathbf{I}_k
\end{array}
\right)
\label{K2}
\end{eqnarray}
on Hilbert space $\mathbb{H}'=\mathbb{H}_a\oplus \mathbb{H}_k$, where $\mathbb{H}_k$ is a $k$-dimensional axillary space. It is easy to prove $\rho \succeq \hat{\rho}$ from   $\rho \succeq p\rho'$ when $k$ is large enough. From Theorem 3, there exists a unitary $W$ and axillary space $\mathbb{H}_{R_1}$ with $d+K$ dimension such that
\begin{eqnarray}
&&{\rm tr}_{r_1}[W(\rho \otimes \mathbf{I}_{d+k})W^\dag]=\hat{\rho}
\nonumber\\
&&{\rm tr}_{a}[W(\rho\otimes \mathbf{I}_{d+k})W^\dag]=\mathbf{I}_{d+k}
\label{K3}
\end{eqnarray}
Now, consider a projection $P$ which projects the state $\hat{\rho}$ onto the subspace $\mathbb{H}_a$, i.e., $\rho'=\frac{1}{p}P\hat{\rho}P$ with success probability $p$. Now, we consider a dephasing quantum channel ${\cal D}_J$ on the subspace $\mathbb{H}_a$ as
\begin{eqnarray}
{\cal D}_J(\hat{\rho})&=&\sum_{i=1}^d\langle i|\hat{\rho}|i\rangle|i\rangle\langle i|
\nonumber\\
&=&\rho'
\label{K4}
\end{eqnarray}
Note that for any dephasing operation \cite{Boes19}, there is a unitary transformation $O=\sum_i|i\rangle\langle i|\otimes V_i$ on the space $\mathbb{H}_a\otimes \mathbb{H}_{r_2}$ satisfying that
\begin{eqnarray}
&&{\rm tr}_{r_2}[O(\hat{\rho}\otimes \mathbf{I}_d)O^\dag]=\hat{\rho}
\nonumber\\
&&{\rm tr}_a[O(\hat{\rho}\otimes \mathbf{I}_d)O^\dag]=\mathbf{I}_d
\label{K5}
\end{eqnarray}
where $V_i$ are unitary operations satisfying ${\rm tr}[V_iV_j]=d\delta_{ij}$, and $\mathbb{H}_{r_2}$ is an axillary space. Define
\begin{eqnarray}
&&U=(O\otimes \mathbbm{1}_{r_1})(P\otimes \mathbbm{1}_{r_1}\otimes \mathbbm{1}_{r_2})(W\otimes \mathbbm{1}_{r_2}),
\nonumber\\
&&\mathbf{I}_{r_1r_2}=\mathbf{I}_{d+k}\otimes\mathbf{I}_d
\label{K6}
\end{eqnarray}
From Eqs.(\ref{K4})-(\ref{K6}), we have
\begin{eqnarray}
&&{\rm tr}_{r_2r_2}[O(\hat{\rho}\otimes \mathbf{I}_{r_1r_2})O^\dag]=\rho'
\nonumber\\
&&{\rm tr}_a[O(\hat{\rho}\otimes \mathbf{I}_d)O^\dag]=\mathbf{I}_{r_1r_2}
\label{K7}
\end{eqnarray}
The success probability is given by ${\rm tr}_{ar_2r_2}[O(\hat{\rho}\otimes \mathbf{I}_{r_1r_2})O^\dag]=p$. This completes the proof. $\Box$

\end{document}